**Accelerating Photovoltaic Materials Development *via* High-Throughput Experiments and Machine-Learning-Assisted Diagnosis**

Shijing Sun[1], Noor T. P. Hartono[1], Zekun D. Ren[1,2], Felipe Oviedo[1], Antonio M. Buscemi[1], Mariya Layurova[1], De Xin Chen[1], Tofunmi Ogunfunmi[1], Janak Thapa[1], Savitha Ramasamy[3], Charles Settens[1], Brian L. DeCost[4], Aaron Gilad Kusne[4], Zhe Liu[1], Siyu I. P. Tian[1,2], I. Marius Peters[1], Juan-Pablo Correa-Baena[1], Tonio Buonassisi[1,2*]

[1]Photovaltaic Research Laboratory, Massachusetts Institute of Technology, Cambridge, MA 02139, USA
[2]Singapore-MIT Alliance for Research and Technology, 138602, Singapore
[3]Institute of Infocomm Research, A*STAR, 138632, Singapore
[4]Materials Measurement Science Division, National Institute of Standards and Technology, Gaithersburg, MD 20899, USA

**Abstract**
Accelerating the experimental cycle for new materials development is vital for addressing the grand energy challenges of the 21st century. We fabricate and characterize 75 unique halide perovskite-inspired solution-based thin-film materials within a two-month period, with 87% exhibiting band gaps between 1.2 eV and 2.4 eV that are of interest for energy-harvesting applications. This increased throughput is enabled by streamlining experimental workflows, developing a set of precursors amenable to high-throughput synthesis, and developing machine-learning assisted diagnosis. We utilize a deep neural network to classify compounds based on experimental X-ray diffraction data into 0D, 2D, and 3D structures more than 10 times faster than human analysis and with 90% accuracy. We validate our methods using lead-halide perovskites and extend the application to novel lead-free compositions. The wider synthesis window and faster cycle of learning enables three noteworthy scientific findings: (1) we realize four inorganic layered perovskites, $A_3B_2Br_9$ ($A$ = Cs, Rb; $B$ = Bi, Sb) in thin-film form via one-step liquid deposition; (2) we report a multi-site lead-free alloy series that was not previously described in literature, $Cs_3(Bi_{1-x}Sb_x)_2(I_{1-x}Br_x)_9$; and (3) we reveal the effect on bandgap (reduction to <2 eV) and structure upon simultaneous alloying on the $B$-site and $X$-site of $Cs_3Bi_2I_9$ with Sb and Br. This study demonstrates that combining an accelerated experimental cycle of learning and machine-learning based diagnosis represents an important step toward realizing fully-automated laboratories for materials discovery and development.

## Introduction

Despite sustained community interest in perovskite-inspired solar cells,[1] progress realizing a suitable lead-free material has been slow.[2,3] The combination of high-throughput experiments and machine learning offers a new approach to explore the rich physics of these materials in device-relevant thin-film form.[4] Herein, we access a wide range of lead-halide and lead-free perovskites within a singular liquid-synthesis thin-film growth environment, by assembling a set of Pb, Sn, Bi, Sb, Ag, Cu, and Na-rich precursors and anti-solvent (chlorobenzene). In a two-month period, we create 96 unique precursor solutions (synthesized using 28 solid precursors based on recently-developed Pb, Sn, Bi, Sb, Cu and Ag compounds, see Table S1 [5–7]), deposit 75 unique chemical compositions into crystalline thin films, and measure 65 materials with bandgaps between 1.2 and 2.4 eV (assuming direct bandgaps), a range auspicious for energy-harvesting applications, *e.g.*, single-junction and tandem photovoltaic (PV) devices.[8] This increased rate of synthesis motivates the development of high-throughput, automated characterization approaches.[9] We acquire X-ray diffraction (XRD) data in ten minutes or less using a high scan rate over a $\theta$-$2\theta$ geometry and use a 3-layer (256-256-256) dense neural network to analyze the structural dimensionality of experimental data with 90% accuracy. The network is trained partially with augmented data from the Inorganic Crystallographic Structure Database (ICSD)[10] to overcome the sparse data problem due to the small experimental dataset. The combination of a wide range of precursors and solvent engineering significantly enlarges the chemical space explored, providing a platform to study the composition map with both single and dual-site alloying. Materials and their alloys targeted and synthesized in this study span $ABX_3$, $A_3B_2X_9$, $ABX_4$, and $A_2B^{I}B^{III}X_6$ ($A$ = MA, FA, Cs, Rb, K, Na; $B$ = Pb, Sn, Ag, Cu, Na, Bi, Sb; $X$ = Cl, Br, I) perovskite-inspired materials with 0D, 2D or 3D crystallographic structures.

Our dual focus on throughput and quality is motivated by the growing gap between theory and experiment. Over one thousand perovskite-inspired candidate compounds have been theoretically predicted during the last few years;[11–13] a triumph of modern first-principles density-functional theory (DFT) simulations, high-performance computing hardware, and the resulting searchable databases of materials properties.[14,15]. Many studies proposed to replace the toxic $Pb^{2+}$ in methylammonium lead iodide (MAPI) and the unstable $Sn^{2+}$ in $MASnI_3$ with $Bi^{3+}/Sb^{3+}$, forming zero- or two-dimensional $A_3B_2X_9$ compounds, or three-dimensional $A_2B^{I}B^{III}X_6$ double perovskites.[16,17] In contrast to the high throughput of theoretical predictions, only a small fraction of predicted compounds has been experimentally realized.[18] A few dozen Pb-free MAPI-inspired compounds were explored, often first by relatively lower-throughput bulk-crystal synthesis.[3,19,20] Among those, Sn perovskites have achieved over 9% efficiency,[21] while the air-stable iodide-based $A_3B_2I_9$ ($A$ = Cs, Rb, MA, $NH_4$; $B$ = Bi, Sb) and bromide-based double perovskite $Cs_2AgBiBr_6$ have recently been incorporated into photovoltaic devices with champion efficiencies over 2%.[16,22,23]

Fast exploration of novel thin-film compositions encounters many practical obstacles, in part because of the complex thermodynamic phase formation of ternary and quaternary compounds, in addition to the unknown kinetics.[5,24] It is time-consuming to access a wide range of the periodic table and possible process windows *via* bulk crystal growth. On the other hand, transforming bulk crystals to the device-relevant thin-film form can be challenging, since *p*-block cations typically have low solubilities in common solvents and process windows are typically less tunable. This often leads to a slow learning rate that favors incremental trial and error[9] and that is at odds with the timelines of many investors and the residency time of individual researchers.[25] As a result, many key breakthroughs, including MAPI solar cells, were discovered rather than predicted.[26] The "experimental bottleneck" limits novel material development both directly and indirectly, *e.g.*, by slowing experimental feedback to improve the predictive accuracy of theoretical models, reducing the success of the inverse-design paradigm.

By increasing experimental throughput without sacrificing material quality, we realize the following scientific contributions in this study: we report four $A_3B_2Br_9$ layered perovskites and their eighteen alloys, which doubles the number of known inorganic bromide perovskite thin-films formed by one-step liquid synthesis (without re-dissolving the product crystals) in the literature.[15] We present the first Pb-free dual-site alloy series, $Cs_3(Bi_{1-x}Sb_x)_2(I_{1-x}Br_x)_9$, which exhibits a transition between 0D to 2D crystal structures and non-linear bandgap tunability. The discovery of anomalous bandgap behavior in this series opens a new pathway to achieve lead-free all-inorganic perovskites for multi-junction solar cells.

In the next sections, we describe how we narrow the "throughput gap" between experiment and theory in three steps: First, we perform each step by hand, quantify our workflow using a timer, and then optimize our workflow to reduce inefficiencies.[27,28] Second, we develop a synthesis platform capable of accessing a wide range of precursors, to evaluate multiple classes of material within an equivalent time. Third, we adapt and harness statistical and machine-learning tools to accelerate data acquisition and analysis.[27,28] We believe that these three steps are broadly applicable to other functional and structural materials searches, given similar reported mismatches between the time needed for each theory and experimental cycle of learning. Notably, these improvements result in a 35× throughput improvement over our laboratory baseline even before implementing systematic and automated solutions; further gains of 10–100× can be expected by fully automating our workflow.

## I. **Workflow Optimization and Precursor Development for a Robust, Flexible Synthesis Platform**

We undertook a multi-year workflow-optimization effort in our laboratory, designed to quantify and reduce the time taken for each step in thin-film synthesis and diagnosis. The goal of our workflow-optimization effort was to maximize "usable information per unit time." We set a target of 35 high-quality compounds synthesized and analyzed per month (both phase and bandgap diagnosis, Figure 1), and targeted a batch throughput of 150 samples per month in our synthesis loop. This throughput is 35× faster than our previous laboratory baselines of PV device fabrication based on other deposition methods,[29–31] approaching the throughputs of HTE platforms developed for other materials classes.[4,27,28,32] See supplemental information for workflow quantification and improvement over time. This allowed an experimental capacity of one cycle of learning per batch per day (Figure 1), including optical and structural characterization, with several candidate materials per batch.

Such fast accumulation of experimental data motivates the development of automated data analysis. Our workflow analysis identified "structural characterization" as an experimental bottleneck limiting the learning rate per sample (see Figure S1 for time breakdown). We then adapt and apply a neural network to assist with phase identification, which enables us to classify and test the crystallographic dimensionality of perovskite-inspired materials within two minutes including the time for training, and within a second after the training in this study. In the context of halide perovskites, the dimensionality of a compound refers to the connectivity of the inorganic framework, *e.g.*, a 3D perovskite, $MAPbI_3$, consists of corner-sharing $PbI_6$ octahedra extended into a 3D network. "Dimensionality" is chosen as a crystal descriptor (*i.e.*, output parameter of merit) in this study, because of recent reports linking charge-transport behavior with higher dimensionality, whereas low-dimensional perovskites tend to exhibit higher stability.[2,33–38] In this study, the dimensionality labels are determined based on the space group and elemental information in the ICSD. Because the amount of labeled data from experimentation alone is too small to adequately train a neural network, we augment experimental training data with powder X-ray diffraction (PXRD) patterns computed from the ICSD. As discussed in the next section, in addition to classification, the neural network can also diagnose the dimensionality of novel materials (Figure 4). For some compounds previously reported in literature, both bandgap and XRD patterns were cross checked with literature data.[39–42]

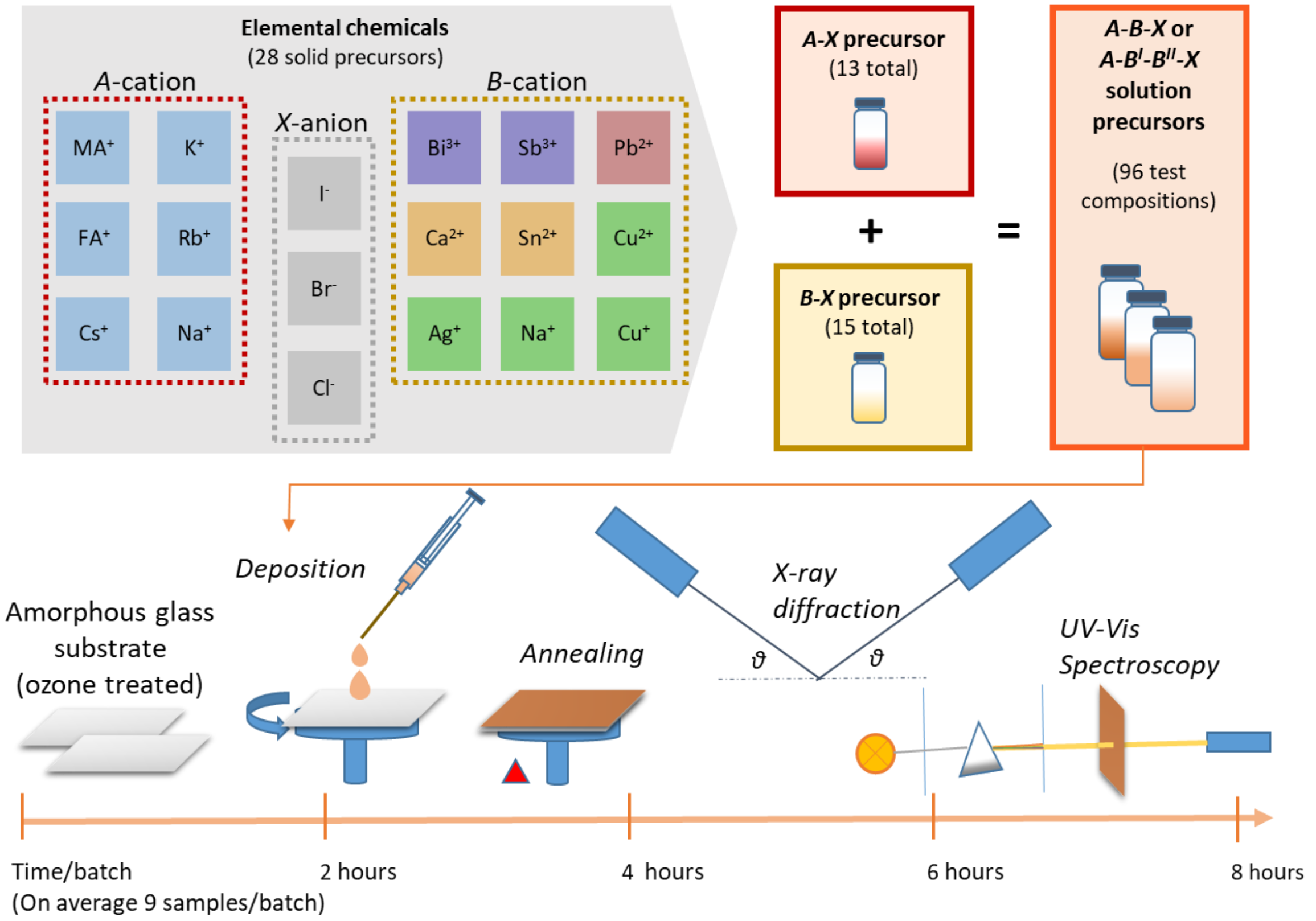

***Figure 1*** *Sketch of the optimized experimental workflow employed in this study. A typical perovskite is a ternary compound with* $ABX_3$ *architecture. Precursor solutions of 96 perovskite-inspired target compounds (test compositions) were prepared by mixing the commercially available A-X and B-X solid precursors in stoichiometric ratios (see supplementary information for materials list and references). Six cations from the periodic table were dissolved as A-site cations in the target compounds, together with nine cations for B-site and three halide ions for X-site. 13 A-X (e.g. CsI) and 15 B-X (e.g.* $BiI_3$*) solid precursors in total were used in this study). Through one cycle of learning, each successfully prepared precursor solution followed the three experimental steps of thin-film deposition, X-ray diffraction and UV-Visible (UV-Vis) spectroscopy measurement to examine the structure and optical properties. Candidate thin-films were screened based on their bandgap and formation of crystalline phases.*

To test a wide composition space while avoiding false negatives, a robust yet flexible synthesis platform is essential. Solution synthesis (*e.g.*, spin coating) of perovskite-inspired materials is fast, but requires liquid precursors amenable to spin coating.[43] A typical perovskite material used in recent solar cells consists of an $ABX_3$ architecture where *A* is a monovalent cation, *B* is lead and *X* is a halide. Historically, the range of precursors was limited, given the low solubilities of *p*-block cations and alkaline earth metals in common solvents. Recently, our laboratory developed Sb and Bi precursors.[44,45] We used these precursors, as well as the commonplace Pb and Sn,[46][47] and the novel Ag and Cu precursors recently reported in literature,[3,48] to centralize our stock solution for synthesis and streamline the parameter space of processing conditions with solvent engineering. We hereto select a list of 96 perovskite-inspired target compounds and alloys for this study by a combination of literature,[6,45,48–52] theory,[12,14] experience, intuition, and in-house research (Tables S1 and S4). Using these test compositions we attempted to synthesize the recently computationally predicted or experimentally realized material classes in thin-film form,[11,14] including the following five perovskite-related crystal systems: $ABX_3$(3D), $A_3B_2X_9$ (2D), $A_3B_2X_9$ (0D), $ABX_4$ (2D), and the 3D double perovskite $A_2B^{I}B^{II}X_6$ ($A$ = MA, FA, Cs, Rb, K, Na; $B$ = Pb, Sn, Ag, Cu, Na, Bi, Sb, $X$ = Cl, Br, I). Precursor solutions were prepared with the stoichiometric ratios of the elements according to the target compounds respectively (Figure 1). Universal solubility test and thin-film deposition methods were applied to access the solution processability of each material in the common organic solvents N,N-Dimethylformamide (DMF) and dimethyl sulfoxide (DMSO). To ensure reproducibility, we synthesized an average of three films per composition. Repetitive synthesis of the key material series is listed in Table S6, where the results were used to train the machine-learning algorithm.

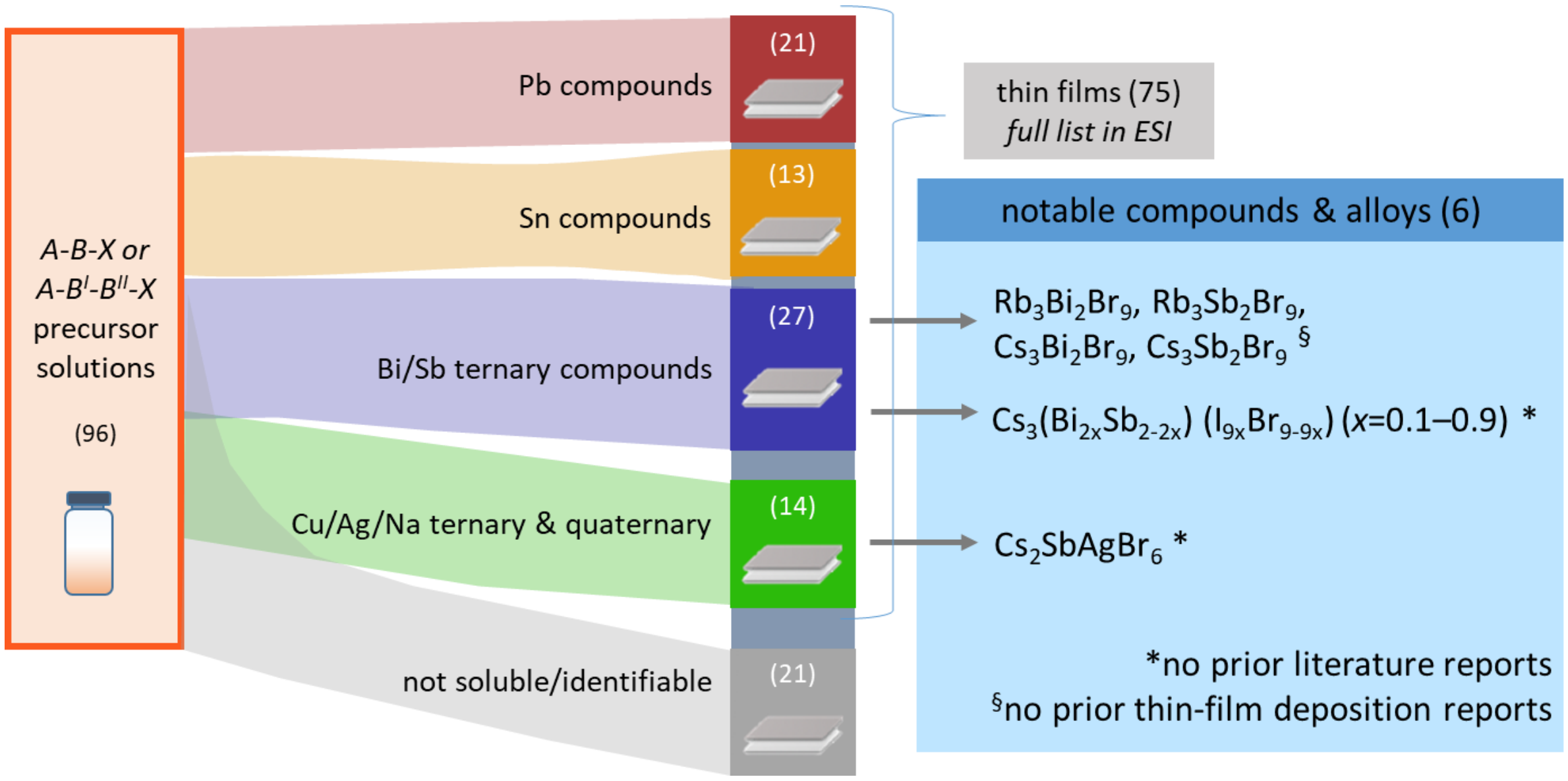

***Figure 2*** *Sankey diagram demonstrating the materials synthesis flow of the 96 precursor compositions attempted, of which 75 resulted in compact thin films. A full materials list appears in the Table S2 and S3 in supplemental information, Repetitive measures including the unsuccessful synthesis reactions for machine-learning training purposes are listed in Table S6.*

We focus our study on four classes of $B$-site cation (Figure 2): Pb-based $ABX_3$ single perovskites; Sn-based $ABX_3$ perovskite alloys, Bi- and Sb-rich $A_3B_2X_9$ low-dimensional perovskite-inspired materials, and Cu-, Ag- and Na- rich perovskite-inspired materials. These four classes represent the well-established thin-film materials (Pb- and Sn-rich), materials explored mostly in the bulk forms (Bi- and Sb-rich), and lesser explored compositions (Cu-, Ag-, and Na-rich), respectively. As shown in Figure 2, 75 out of the 96 attempted test compositions successfully yielded crystalline thin-films. Among these 75, 21 are Pb-halide perovskites with dual-site alloying: $MA_{1-x}FA_xPbI_xBr_{3-x}$ and $(Cs_{0.05}Rb_{0.05}MA_{0.9-x}FA_x)PbI_xBr_{3-x}$, which were deposited following literature deposition methods as a reference for our workflow timeline and verification for the machine-learning techniques;[7,39,49] 13 were Sn-rich perovskites with single and dual-site alloying: $MASn_{1-x}Ca_xI_3$ and $MASn_{1-x}Pb_xI_{3-x}Br_x$ in which we investigated the trend of bandgaps with multi-site dopants, as discussed in the next paragraph; 27 are Bi- and Sb-based ternary compounds with single and dual-site alloying, including $A_3B_2X_9$ ($A$=MA, Cs, Rb; $B$ = Bi, Sb; $X$ = I, Br), $Cs_3Bi_2Sb_{2-2x}I_9$, and $Cs_3(Bi_{1-x}Sb_x)_2(I_{1-x}Br_x)_9$; and 14 include incorporation of Ag, Cu and Na in the precursor with perovskite-like stoichiometry of $MA_2CuX_4$ ($X$ = Cl, Br) and $A_2B^IB^{II}X_6$ double perovskites ($A$ = Cs, Rb; $B^I$ = Bi, Sb; $B^{II}$ =Ag, Cu, Na; $X$ = I, Br). In addition to the 75 thin films, we discarded 21 precursor solutions at the thin-film deposition stage, most of which belonged to the Cu/Ag/Na ternary and quaternary compound class. The unsuccessful syntheses were attributed to the low solubility of the reactants in the solution (*e.g.* cesium acetate with silver bromide) or poor film formation within processing window (e.g. $Na_3B_2X_9$ ($B$ = Sb, Bi)).

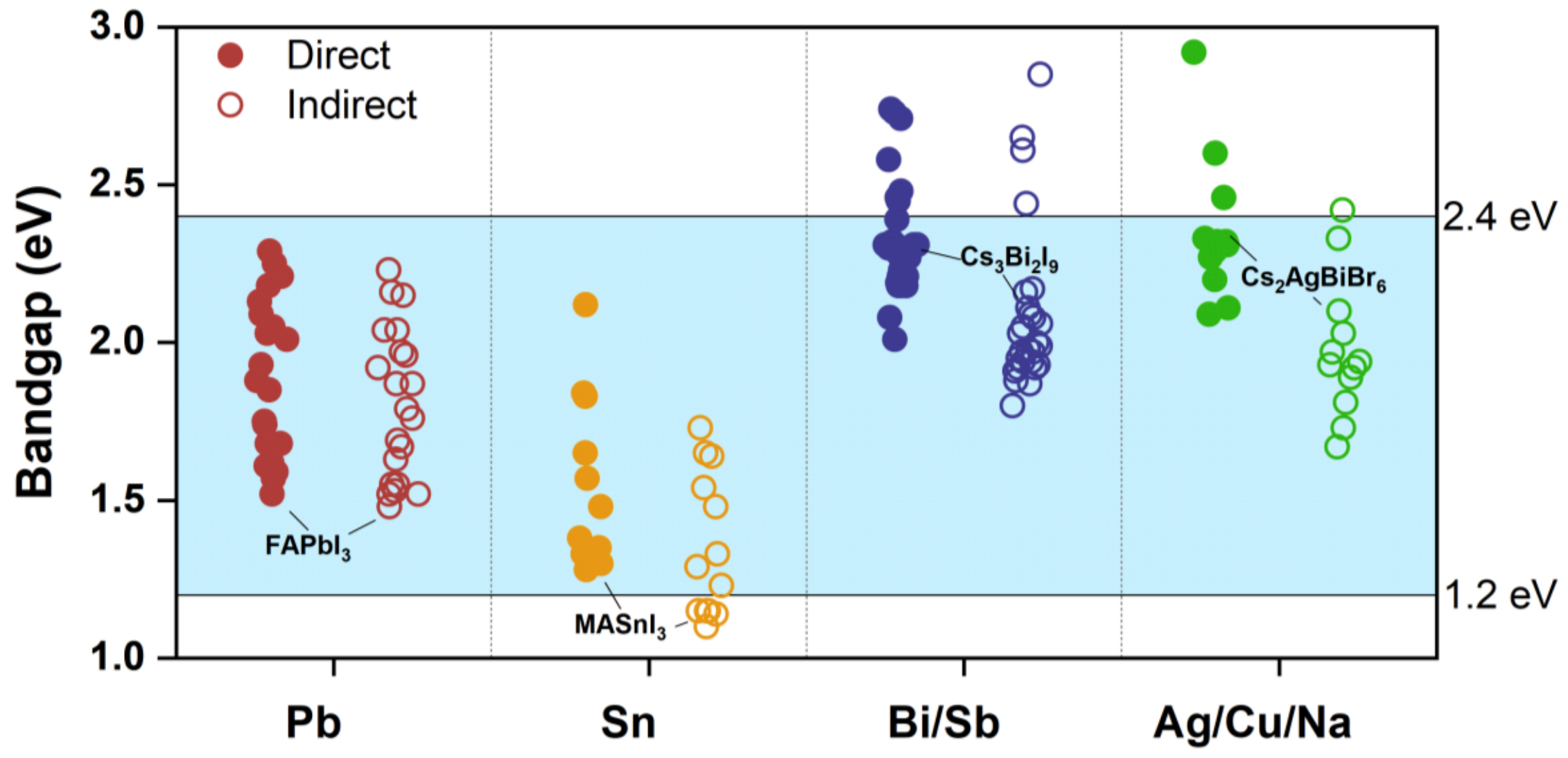

***Figure 3*** *Bandgaps measured for the 75 thin-film materials investigated in this study. Bandgaps were extracted from Tauc plots (Figure S7) assuming direct (solid dots) and indirect bandgaps (circles). Assuming direct bandgaps, 65 of the 75 materials show bandgaps between 1.2 and 2.4 eV, of strong interest for energy-harvesting applications.*

Figure 3 presents the optical properties of the 75 materials grouped by $B$-site compositions We measured the reflection and transmission of each film and bandgaps were deduced from Tauc plots (Figure S7).[53] Both direct and indirect bandgaps were calculated from measurements of freshly made films, to avoid prior bias on the unknown crystal structures and any degradation. It is well established that introduction of Br on the $X$-site of $MAPbI_3$ increases the bandgap from 1.5 (0% Br) to 2.35 eV (100% Br).[54] The results from our measurements confirm this trend with mixed cation and halide alloys. On the other hand, when both Br and Sn are alloyed into $MAPbI_3$ simultaneously, our results indicate that adding $PbBr_2$ into the $MASnI_3$ system increases the bandgap, suggesting the effect of Br substitution for I has a more dominant effect than Pb substitution for Sn, an effect one might assume to be influenced by relative differences in electronegativity between substituting atomic pairs.[47] Furthermore, we tested 62 (out of 96) Pb-free and Sn-free target compounds and alloys in this study and 41 of them yielded in compact films. We expanded the known inorganic Bi and Sb halide perovskite deposition methods from iodide to bromide, the latter having received much less attention, and successfully developed $A_3B_2Br_9$ ($A$ = Cs, Rb; $B$ = Bi, Sb) thin films. This further enabled us to explore a range of multi-site alloys within the Bi/Sb material classes and achieved bandgap tuning between 1.2 and 2.4 eV via compositional engineering as shown in Figure 3 and Table S2.

**II. Machine-learning assisted Structural Diagnostics**

A neural network is applied in this study to extract the materials descriptor, "dimensionality", from hundreds of experimental PXRD patterns. Upon manual phase identification,[41,55,56] 55 out of the 75 thin-film materials are first identified with seven space groups (Table S2), belonging to three crystallographic dimensionalities where 11 are 0D (molecular dimers, *e.g.*, $Cs_3Bi_2I_9$), 10 are 2D (layered perovskites, e.g. $Cs_3Sb_2Br_9$) and 34 are 3D (conventional perovskites, *e.g.*, $MASnI_3$ and $CsAgBiBr_6$). While it is promising to replace future human synthesis with robotics,[18] we soon realized that the traditional methods of analyzing XRD data are far slower than the rate of high-throughput film deposition and require significant human expertise along with prior knowledge of the expected crystal structures. Our accelerated material development paradigm tackles this issue by implementing a deep neural network classifier based on X-ray diffraction peak positions, allowing a rapid experimental screening of candidate materials based on structural features. We make a detailed comparison of various machine learning techniques and data augmentation algorithms in our parallel study on method development[57] showing the advantages in speed and accuracy of the deep neural network presented in this paper.

In order to learn effective XRD representations in the context of our small (<100 samples) dataset, we applied a transfer learning approach. A training dataset was first built consisting of 164 PXRD patterns, and their structural labels were extracted from ICSD. Single, double, ternary and quaternary combinations of the elements and space groups of interest were covered in this data-mining, featuring the target 0D, 2D, and 3D perovskite materials and potential byproducts (precursors). To better account for the difference between the simulated powder patterns from the crystallography database and the experimental thin-film diffraction patterns, both simulated patterns and experimental patterns were subjected to a data augmentation process of sequential random peak scaling/elimination and peak position shifting to account, respectively, for the preferred orientation and substrate-induced strain in the thin-film samples. After the data augmentation process, 50,000 XRD patterns with corresponding dimensionality labels were obtained.

We first use this dataset to train a deep feedforward neural network, consisting of three hidden layers with *ReLU* activation functions of 256 neurons each.[58,59] The classification layer uses a *softmax* activation function, so that the neural network outputs can be interpreted as probabilities for each potential dimensionality. The weight optimization to minimize a log-loss function was performed by stochastic gradient descent (constant learning rate of 0.01)[60]. L2 regularization with a $\lambda$ of 0.0001 was used, and the number of training epochs was set to 400. The dimensionality classes of the training dataset were correctly balanced during database mining to avoid issues caused by class imbalance. The accuracy of the test model is defined as the number of correctly labeled XRD patterns among the whole testing dataset, independently of class, compared to the ground-truth labels in ICSD. In this first approach, a mean model accuracy of 99% was achieved using 5-fold cross validation on the simulated XRD patterns and their structural labels.

The second approach consisted of using the simulated XRD patterns as a training dataset, and the experimental patterns as a testing dataset using the same neural network architecture and hyperparameters. A mean model accuracy of 76% was obtained for 5-fold cross validation. Although the data augmentation provided an improvement compared to a baseline classification accuracy of less than 60% when non-augmented data is used for training, using only simulated XRD patterns for training seems to capture all the subtle differences among experimental XRD spectra, such as systematic experimental error in sample alignment and random human errors in synthesis. To accurately capture these experimental features, the collected PXRD patterns for the 75 unique compositions were subdivided into two experimental datasets: *group 1* - manually identified perovskites with space group and dimensionality labels (55); and *group 2* - new materials without structure details reported (19). Figure 4(a) represents the schematic illustration of the third approach, where the final training dataset consists of simulated and randomly selected 80% of the *group 1* experimental patterns. Subsequently, after cross validation, the blindfold model accuracy of 90% was achieved for *group 1* materials. Significant improvement in the model accuracy and robustness was demonstrated using the experimental data as part of the training set.

Figure 4 (b) depicts the classification of the 75 unique compositions into three crystallographic dimensionalities based on the third machine learning approach, where a confidence score between 0 and 1 was assigned to 0D, 2D and 3D based on the *softmax* activation function outputs from the neural network, respectively for each composition. The *softmax* activation function assigns decimal probabilities to each class in a multi-class problem and a confidence score of > 0.5 indicates the suggested dimensionality of a given material.[59] Most data points in Figure 4 (b) cluster at the vertex of the triangle; the neural network classifier successfully separated the materials with different crystallographic dimensionalities with a relatively high confidence score. Considering the four materials classes synthesized in this study, Pb (yellow data points in Figure 3(b)) and Sn-rich (cyan) perovskites in this study exhibit high confidence scores that point to 3D structures, which are consistent with the structural labels from a large number of literatures on $ABX_3$ ($A$ = MA, FA, $B$ =Pb, Sn, $X$ = I, Br).[14,61] The addition of Ag/Cu/Na in the halide perovskite-inspired materials (navy), on the other hand, results in various mixed phases that are challenging to identify. For example, the composition Ag + CsBiBr shows a confidence score of 0.98 for the 3D dimensionality (which comprises space groups $Pm\overline{3}m$, I4/mcm or $Fm\overline{3}m$), the compound was in fact confirmed to exhibit the same PXRD pattern as the reported 3D double perovskite, $Cs_2AgBiBr_6$.[3] On the other hand, it is likely that the majority phase in the Na + CsSbI recipe we developed in house does not have the same symmetry (3D confidence score of of 0.05). This suggests that the target phase, theoretically predicted double perovskite $Cs_2NaSbI_6$ with $Fm\overline{3}m$ space group was likely not formed under the applied experimental conditions (Table S3). Furthermore, all the synthesized Bi/Sb ternary compounds (red) were classified to non-3D groups, which is consistent with the literature reports of $A_3B_2X_9$ ($A$ = Cs, Rb, MA, $B$ = Bi, Sb, $X$ = I, Br). With these structural guidelines on novel materials, off-line in-depth studies on the synthesis and characterization of the 19 *group 2* compositions are currently under further investigation. The machine-learning method discussed in this section provides a suggested structure type for hundreds of experimental PXRD patterns within seconds, which significantly shortens the diagnostics time, and most importantly, it enables the incorporation of structural characterization into the high-throughput cycle of learning for future laboratory design. With increasing amounts of training data recorded in a future fully automated laboratory, more high confidence score is expected to be achieved, however, samples with low confidence scores still need manual diagnostics to identify the correct structural features. Nevertheless, the introduction of machine-learning techniques to XRD diffraction analysis is also particularly helpful for materials undergo phase transition, where it can be challenging to analyze multiphase problems with manual refinement. The application of this machine-learning approach on a set of *group 2* test materials, $Cs_3(Bi_{1-x}Sb_x)_2(I_{1-x}Br_x)_9$, is shown in the next section, demonstrating a case study to extract the dimensionality for novel $A_3B_2X_9$ alloys based on PXRD patterns.

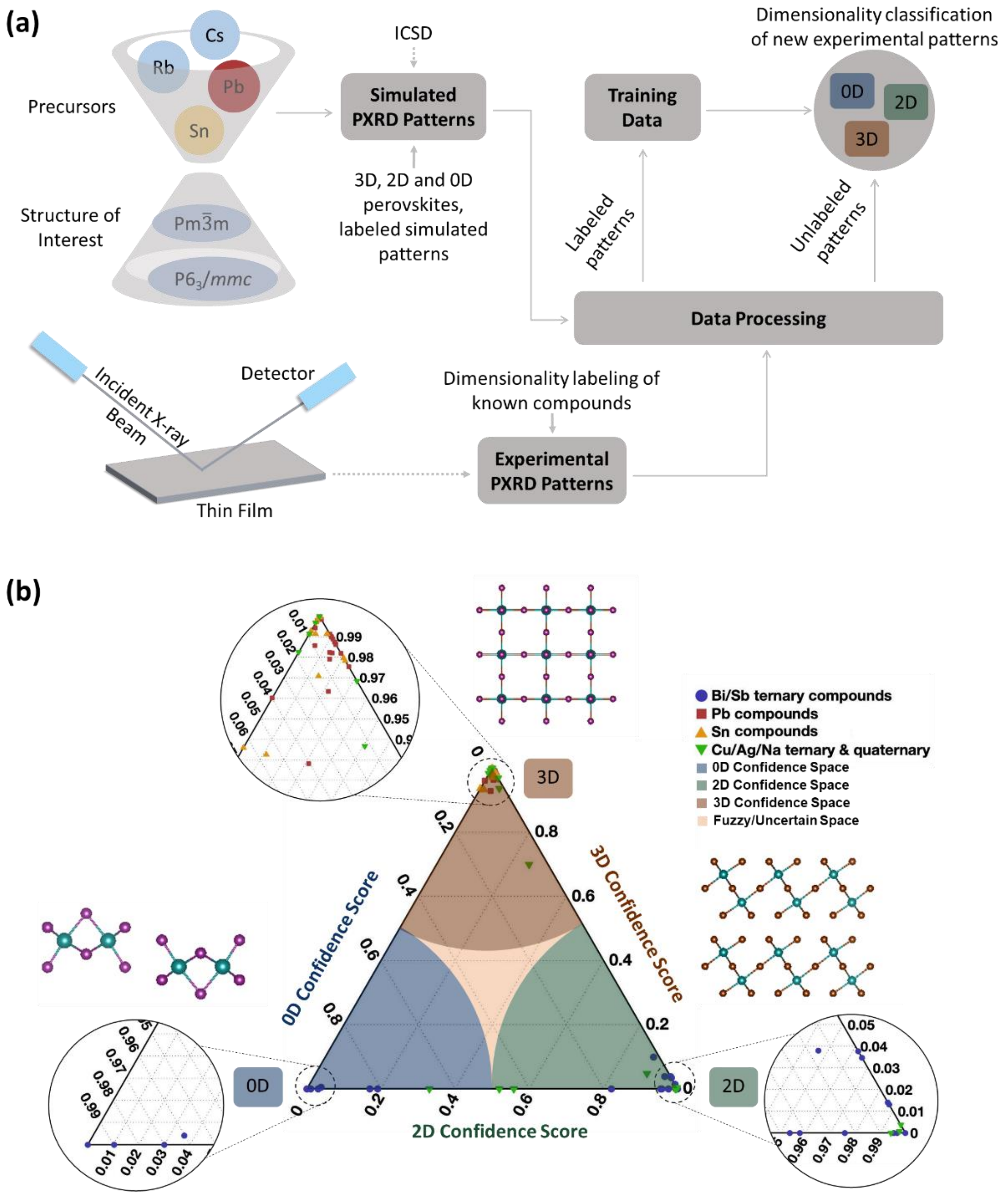

*Figure 4: (a) A schematic workflow to employ machine-learning algorithms to assist structural characterization for perovskite-inspired materials and guide the analysis of diagnosis results. The input osf the algorithm applied is the simulated and experimental PXRD patterns along with a list of precursor elements and expected crystal symmetry for target compounds (based on theory or literature). The output is the crystallographic dimensionality of the experimentally synthesized compounds. (b) An overview of the machine-learning assisted structural diagnostics results. 75 unique compositions were classified into 0D, 2D and 3D perovskites based on the crystal symmetry of the training dataset. A confidence score is associated with each output from the softmax layer in the neural network. The inorganic frameworks of the molecular dimer, layered perovskite, and cubic perovskite crystal structures of interest to this study are drawn next to 0D, 2D and 3D labels respectively.*

**III. New, tunable materials in thin-film form**

With improved throughput and access to a wider range of precursors in a singular growth environment, a wide chemical space is explored. Twenty Br alloys are synthesized, more than double the known thin-film all-inorganic Br-based perovskite-inspired materials in the literature.[3,62] Here, we describe previously unreported halide phases, as well as thin-film versions of 2D layered perovskites that were previously only grown in bulk-crystal form.

First, the signature of a new double perovskite phase, $Cs_2SbAgBr_6$, was detected in the form of a thin-film mixture. We calculate a direct bandgap of 1.89 eV (2.27 eV if fitting assumes indirect bandgap) and the compound is studied in depth in collaboration with Wei *et al.*[63] Second, we succeeded in growing thin films of several perovskite structures that have previously only been reported as bulk crystals. Upon the successful dissolution of RbBr, CsBr, $BiBr_3$ and $SbBr_3$ in various ratios of DMF and DMSO and with annealing temperature control (Table S3), $A_3B_2Br_9$ ($A$ = Cs, Rb; $B$ = Bi, Sb) compounds were successfully realized in compact thin films via one-step solution synthesis, without the need of re-dissolving the product nanocrystals synthesized via bulk synthesis back in precursor solution as demonstrated in previous reports.[64] 2D layered perovskite structures are confirmed for $Rb_3Bi_2Br_9$, $Rb_3Sb_2Br_9$, $Cs_3Sb_2Br_9$ and $Cs_3Bi_2Br_9$.[56] The corner-sharing double layers observed in these halide perovskites typically show a direct bandgap and were reported with higher current than their 0D counterparts in photovoltaic devices.[45]

Third, we observe that the bandgap is not continuous over the transition from 0D to 2D structure in a new alloy series, $Cs_3(Bi_{1-x}Sb_x)_2(I_{1-x}Br_x)_9$ ($x$=0.1–0.9), where the alloy has a lower bandgap than either of the two end phases. The 2D Br-based $A_3B_2Br_9$ perovskites discussed in the previous paragraph are desirable for achieving higher charge transport than their 0D I-based $A_3B_2I_9$ counterparts, however, their bandgap are considered to be too high for promising solar applications (>2 eV). Therefore, new materials that combine higher dimensionality and lower bandgap are desirable. We herein present an important finding resulting from our materials search, demonstrating a dual-site alloy, $Cs_3(Bi_{1-x}Sb_x)_2(I_{1-x}Br_x)_9$ ($x$=0.1–0.9). Figure 5 (a) illustrates the crystal structure of the two end-members of this alloy series, $Cs_3Bi_2I_9$ and $Cs_3Sb_2Br_9$. Pawley refinement of the PXRD patterns of the thin films confirm that the two materials exhibit the 0D $P6_3$/mmc and 2D P-3m1 space groups, respectively (Figure S8). The absorption edge, on the other hand, shifts to higher wavelength before switching to the other direction with more $SbBr_3$ added (Figure 5 (b)). The crystal structure transforms from the 0D $Cs_3Bi_2I_9$ to the 2D $Cs_3Sb_2Br_9$ with increasing $SbBr_3$ content in the precursor solution (Figure 5(c)). Major peak positions shift to the right due to a contraction in lattice parameters. Distinguishing between 0D and 2D $A_3B_2X_9$ compounds has been difficult based on manual phase identification from PXRD measurement, since many of the peak positions overlap and only subtle differences are observed between space groups.[16,65]

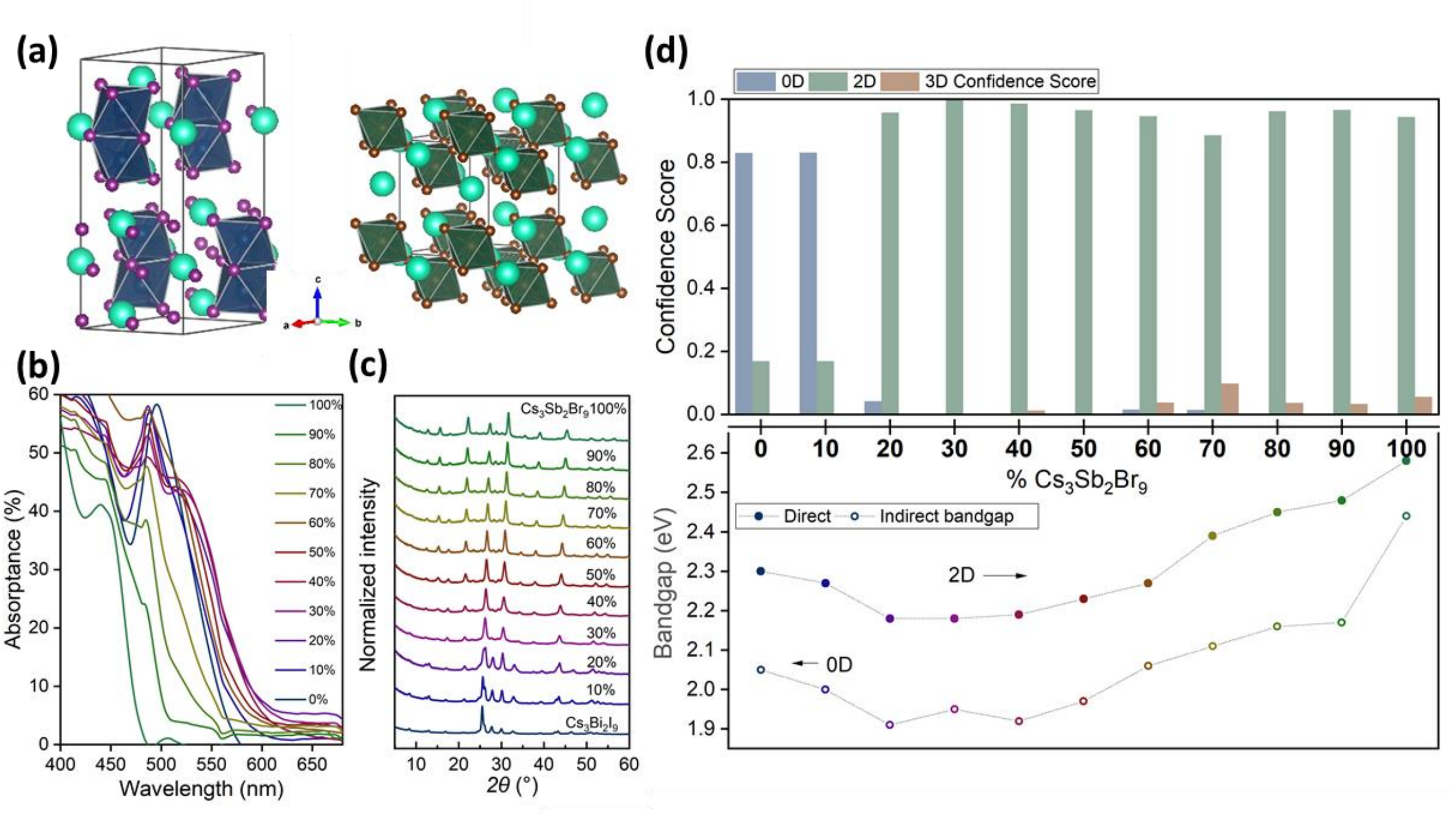

*Figure 5: (a) Crystal structure and PXRD patterns of the 2D $Cs_3Sb_2Br_9$ (top) and the 0D $Cs_3Bi_2I_9$. Pawley Refinement on the PXRD patterns confirm the phase (Figure S7). (b) and (c) X-ray diffraction and absorptance measurement for structural and optical characterization of the of the* $Cs_3(Bi_{1-x}Sb_x)_2(I_{1-x}Br_x)_9$ *($x$=0.1–0.9) respectively. (d) Indication of crystallographic dimensionality based on the machine-learning approach. Bandgaps were calculated accordingly assuming direct and indirect bandgaps. Photographs of this alloy is shown in Figure S6.*

As shown in Figure 5(d), machine-learning-assisted diagnostics indicate that the change of crystallographic dimensionality takes place upon doping 10 - 20% $SbBr_3$ in this experiment. Interestingly, while the material undergoes a structural change, there is also a change in the bandgap, a phenomenon that is observed for the first time in lead-free perovskite-inspired materials. A decrease in bandgap is observed in the 0D region, with increasing Sb and Br content, contrary to expectations that smaller atoms result in tighter binding and larger bandgaps. With 20% $SbBr_3$ doping, the bandgap was reduced to 1.9 eV assuming an indirect bandgap, which is lower than that reported for $Cs_3Bi_2I_9$ and $Cs_3Sb_2I_9$.[44,45] Note that the observation of the bandgap trend in the 20%-doped alloy is not dependent on the assumption of either direct or indirect bandgap during fitting of optical absorptance data. We speculate one possible mechanism for this behavior may be that while the increase in the Br content in the alloys tends to increase the bandgap, the lattice disorder introduced by Sb in the Bi alloys reduces the overall bandgap. The single-site alloy series of $Cs_3(Bi_{1-x}Sb_x)_2I_9$ ($x$=0.1–0.9) also shows a similar "bowing" trend (Table S2). Previous reports on single-site alloys show anomalous bandgap behavior in Pb-Sn solid solution.[66,67] The mechanism of Sb incorporation into the Bi-based perovskite systems is still unclear and is under further investigation. In the 2D alloy region, on the other hand, the increasing Br and Sb concentrations increase the bandgap as expected, as shown in Figure 5 (d).

**Conclusions**:

We here demonstrate a case study on perovskite-inspired materials that the gap between exploration rates of theory and experiment has been closed by one order of magnitude via fast synthesis and machine-learning assisted diagnostics. This framework represents an important step toward a fully-automated lab of the future for discovering functional inorganic and hybrid materials. We utilize a combination of traditional and machine-learning-aided approaches to overcome bottlenecks in materials screening and down-selection, precursor development, workflow optimization, and automation of characterization output. We design and realize a high-throughput experimentation platform capable of investigating 75 unique compounds in two months, using 96 precursor combinations. 87% of the thin films synthesized fell within the bandgap range of 1.2 to 2.4 eV, promising for opto-electronic applications. A neural network was employed to assist in structural analysis, which achieved 90% accuracy in distinguishing the crystal dimensionality of perovskite-inspired materials in this study. This approach is fast, easy to use, and assists chemists to quickly identify, for example, whether 3D perovskites were synthesized during a high-throughput screening.

With this accelerated platform, we realized four lead-free layered perovskites, $A_3B_2Br_9$ ($A$ = Cs, Rb; $B$ = Bi, Sb) and their multi-site inorganic alloy series, $Cs_3(Bi_{1-x}Sb_x)_2(I_{1-x}Br_x)_9$ in compact thin-film form. We examine the “bending” trend in bandgaps in the alloy series and correlated this with a 0D - 2D structural transition in crystallographic dimensionality, which was identified by machine-learning classification. Most importantly, the combination of increased experimental throughput and the successful application of statistical diagnostics provide a new paradigm to examine structure-property relationships, finding non-intuitive trends in a multi-parameter space. Such techniques have been under rapid development in recent years, and will be increasingly easy to access on a daily basis for researchers in the lab.

**Acknowledgements**:
We thank Vera Steinmann and Seongsik Shin for assistance in workflow quantification; and J. Alex Polizzotti, Jeremy P. Poindexter, and Rachel Kurchin for fruitful discussions. Fruitful discussions with Fengxia Wei, Anthony Cheetham and Yue Wu on lead-free perovskite synthesis and diffraction pattern visualization are appreciated. We thank Qianxiao Li (from A*STAR) for inspiring discussions on various machine learning techniques.

This work was supported by a TOTAL SA research grant funded through MITei, US National Science Foundation grant CBET-1605547, and Singapore’s National Research Foundation (NRF) through the Singapore-Massachusetts Institute of Technology Alliance for Research and Technology’s Low Energy Electronic Systems research programme; S.R.’s work was supported by AME Programmatic Fund by the Agency for Science, Technology and Research under Grant No. A1898b0043. The use of the X-ray

Diffraction shared experimental facility at Center for Materials Science and Engineering, MIT was supported by Skoltech as part of the Skoltech NGP Program.

# Supplemental Information

## I. Workflow Quantification and Optimization

Our efforts to quantify and optimize laboratory workflow extended over several years. We quantified the human time per sample for perovskite device fabrications in our lab in the past 3 and half years.

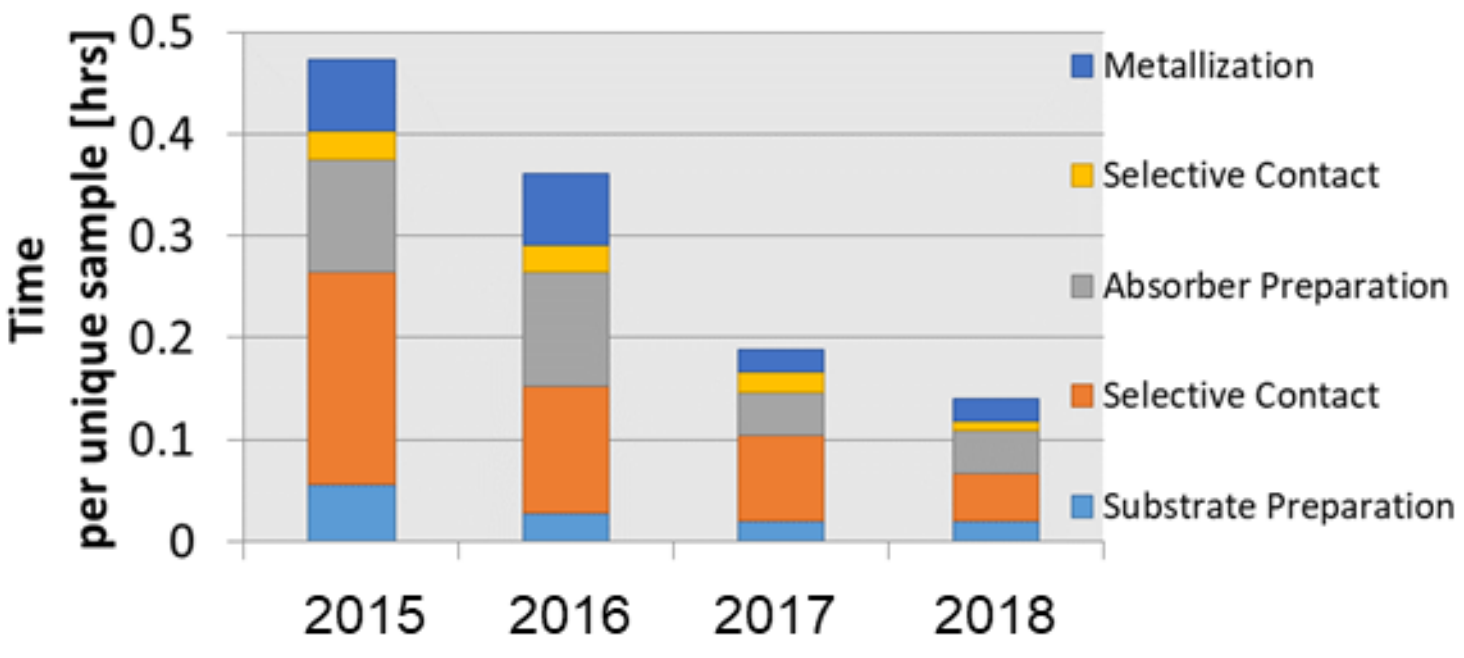

Figure S1 Quantified average time taken for device fabrication at MIT PV Lab over the past 4 years. All data was collected on perovskite solar cells fabrication via solution synthesis. The reduction in time per sample was achieved by the equipment investment that enabled parallel sample fabrication and the workflow optimization of each individual process.

Once the workflow was measured, it could be optimized. Our workflow quantification identified several experimental bottlenecks that required disproportionate time, including sample synthesis and phase identification, which focused our workflow-optimization efforts. We de-bottlenecked our sample-synthesis workflow by selecting a singular synthesis platform with high throughput and flexible, low-cost precursors: solution synthesis by spin coating. We moved all equipment under one roof in a singular material flow, reducing experimental overhead. We de-bottlenecked our analysis workflow by co-optimizing the scan conditions (*e.g.*, resolution) and the machine learning algorithms to analyze the measurement outputs. We were able to reduce the time required to determine dimensionality from X-ray diffraction measurements from several hours to under seven minutes. The next two sections describe our efforts to centralize stock solution synthesis and develop machine-learning-assisted diagnostics, which enabled the >10x faster materials development cycle.

The figure below shows the extension of our current cycle of learning with a process-optimization feedback loop, enabled by AI/ML. In this study, the process optimization (feedback from diagnosis to synthesis and theory) was performed manually. However, at the interface of human and robotic-dominated experimentation, current laboratory workflow has to be adapted, transferring materials knowledge (*e.g.*, precursor chemistry) into an automatic feedback loop.

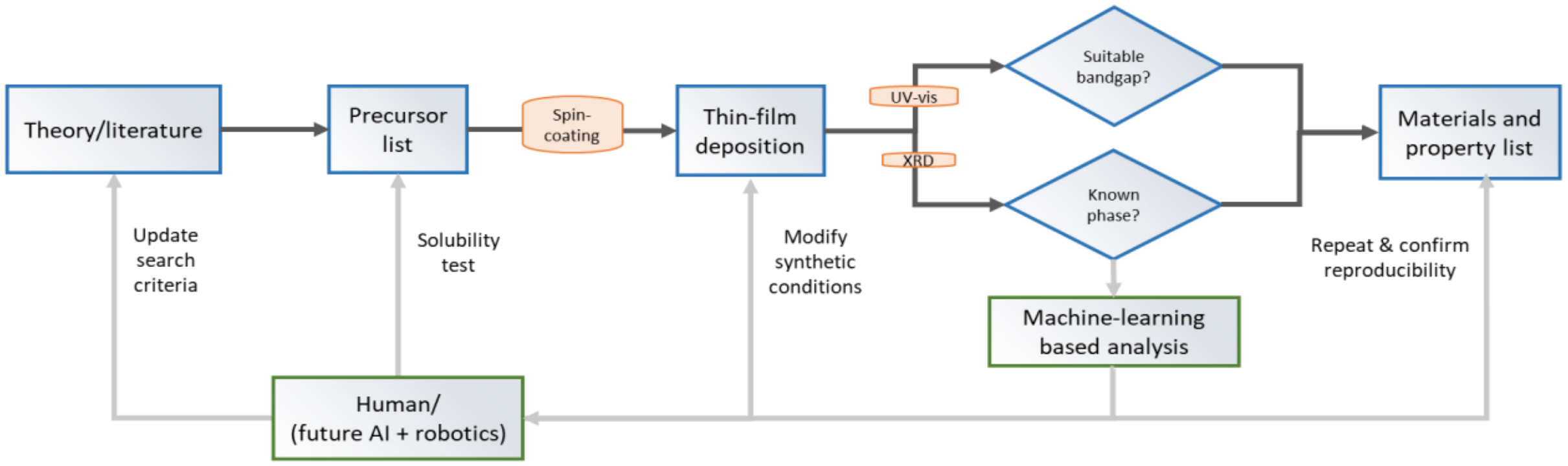

Figure 2 Optimized workflow with the vision for future laboratory incorporating robotics, machine learning and artificial intelligence to replace the human-intensive experiments and diagnostics.

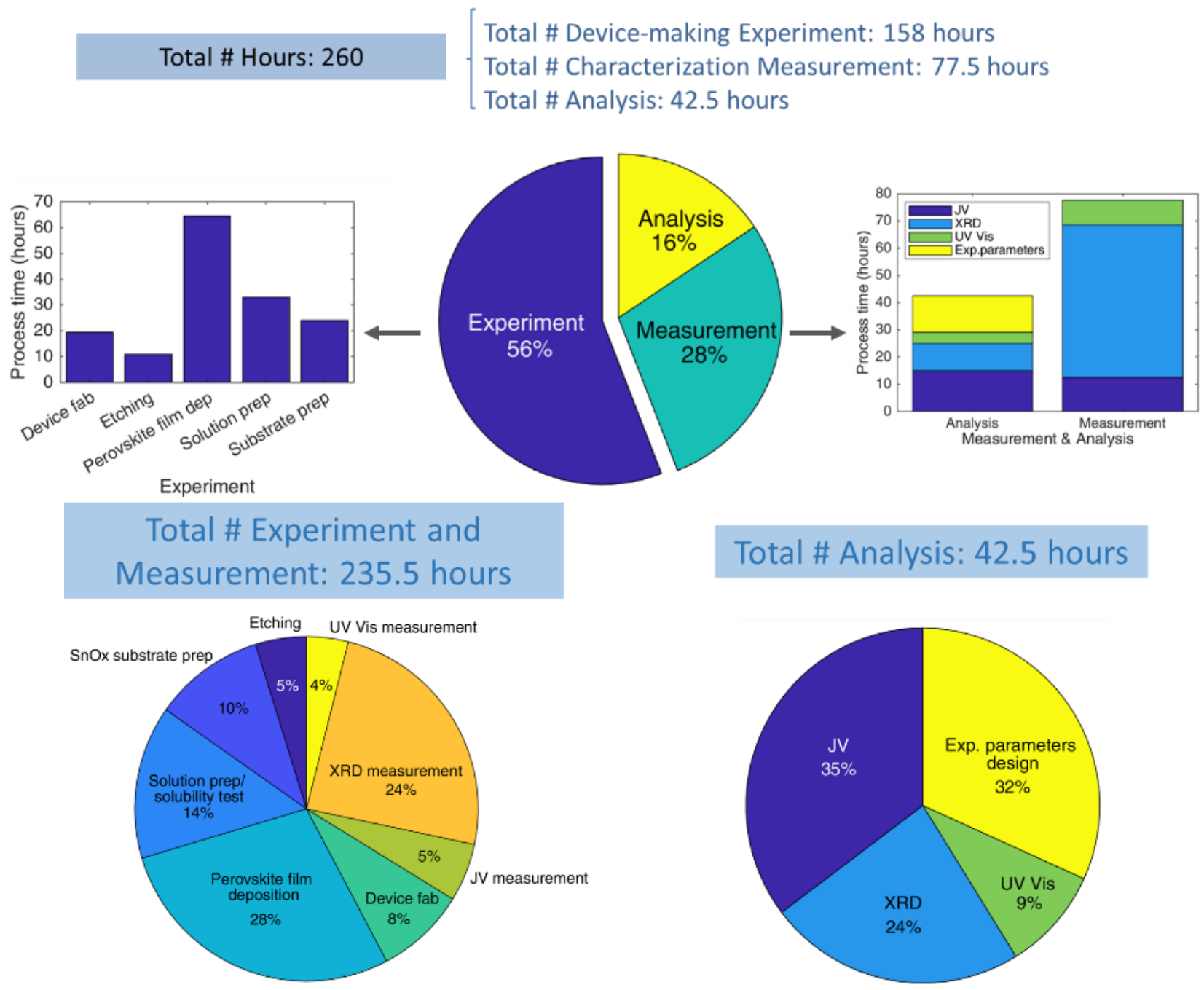

Figure S3 Quantification of the time breakdown per task during a 260 working hours testing period at MIT PV Lab for perovskite solar cell fabrication over the 2 months period of this study. The measurement and analysis of the structure characterization was a rate-determining step in the materials discovery phase, before investment and decision of device fabrication to be continued.

## II. Materials and Methods

Based on the different *B*-site metal cations employed, four classes of perovskite systems were synthesized over our campaign, from the well-established Pb and Sn based 3D perovskites, to the less-understood Bi and Sb perovskites, and exploring new materials based on introducing Ag and Cu into the solvent systems under a singular growth environment.

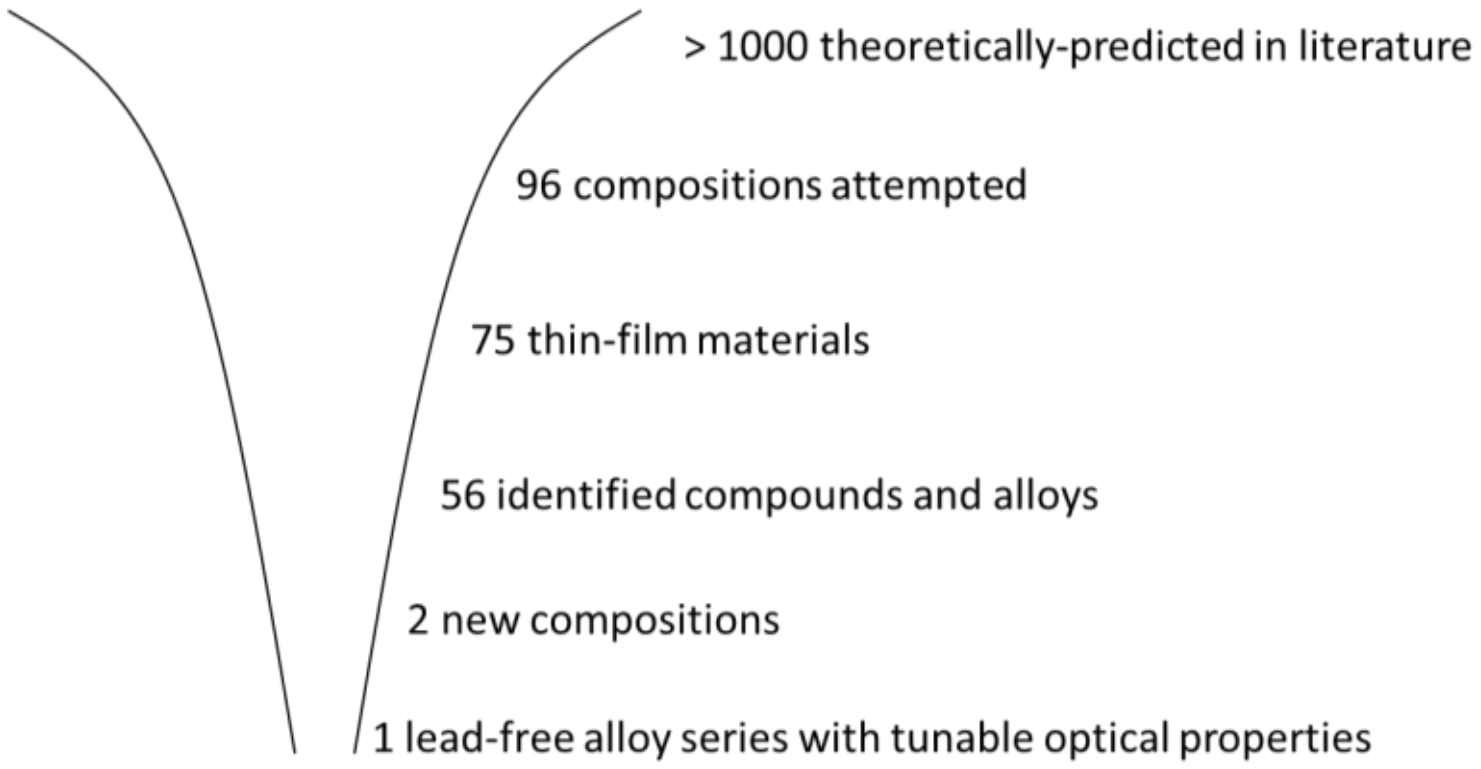

Figure S4 Down-selection of the 96 perovskite-inspired materials that are presented in this study.

Table S1 List of 28 solid precursors used in this study:

| Solid Precursors | | |
|---|---|---|
| A-X | B1-X | B2-X |
| CsI | AgI | $BiI_3$ |
| RbI | AgBr | $SbI_3$ |
| MAI | NaI | $BiBr_3$ |
| MABr | NaBr | $SbBr_3$ |
| MACl | CuI | $BiCl_3$ |
| FAI | $CuCl_2$ | $SbCl_3$ |
| CsBr | | $PbI_2$ |
| CsCl | | $SnI_2$ |
| $CsCH_3COO$ | | $PbBr_2$ |
| KI | | |
| RbBr | | |
| FABr | | |
| CaI | | |

**Pb and Sn-rich ternary compounds:**

Pb and Sn based compounds were synthesized by mixing equal molar solution of AX (A=MA, FA, Cs, Rb, X = Br, I) and $BX_3$ (B = Pb, Sn, Ca, X = Br, I) in solvent of DMF : DMSO = 9:1. An excess $PbI_2$ was added in the $(CsRbMAFA)Pb(IBr)_3$ series (Table S2) to improve the stability following the literature reports.[1] A two-step spin-coating program was employed with 1000 rpm for 10s and then 6000 rpm for 30s. 150 µL of chlorobenzene was added within 2s of the second step as an antisolvent.[2,3] All thin-films deposited in this study was on 1 inch x 1 inch amorphous glass slides.

**Lead-free ternary compounds:**

Bi, Sb and Cu based compounds were synthesized by mixing stoichiometric molar solution of AX (A=MA, Cs, Rb, K, Na, X = Cl, Br, I) and $BX_3$ (B = Bi, Sb, X = Br, I) or $CuCl_2$ in mixed solvents (Table S2). A one-step spin-coating program (2000 rpm for 30s) was used for the first synthesis round. A two-step spin-coating program was then employed with 1000rpm for 10s and then 6000rpm for 30s. 150 μL of chlorobenzene was added within 2s of the second step as an antisolvent.[4,5]

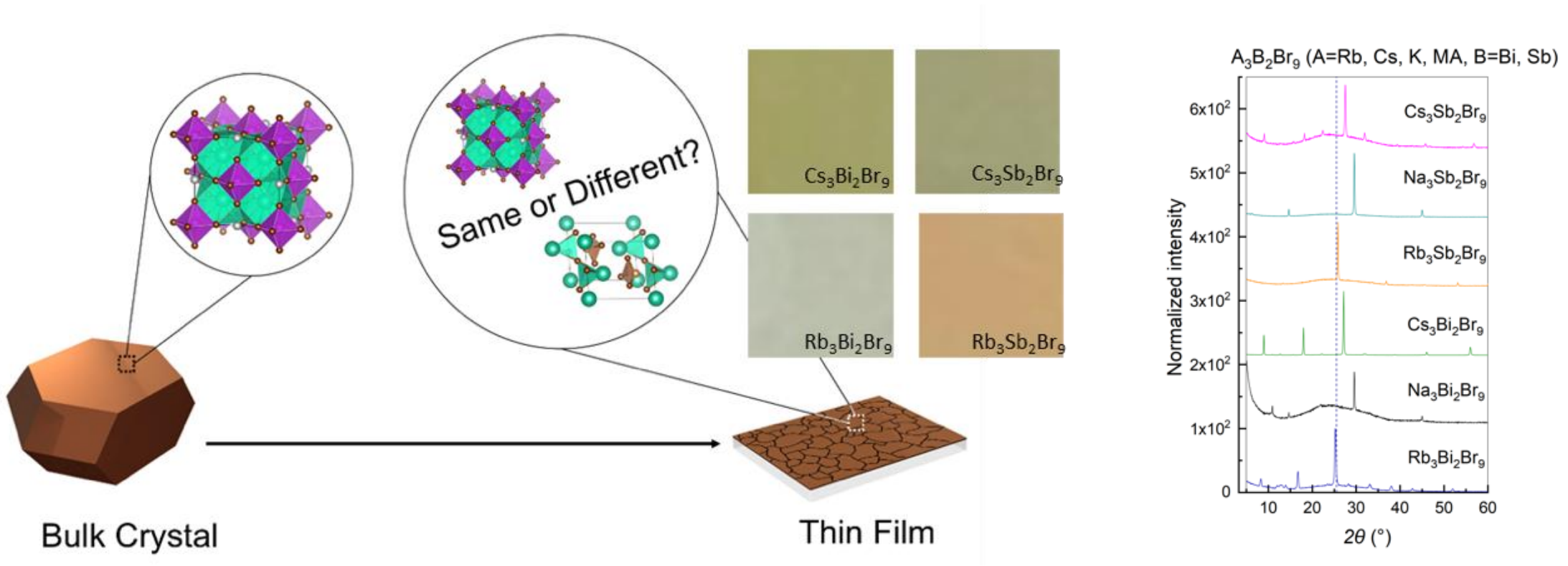

Figure S5: $Cs_3Bi_2Br_9$, $Cs_3SbBr_9$, $Rb_3BiBr_9$, and $Rb_3SbBr_9$ phases are successfully realized in thin-film form. $Na_3Sb_2Br_9$ and $Na_3Bi_2Br_9$ stoichiometry were also deposited, where the thin film phases were not found in ICSD database.

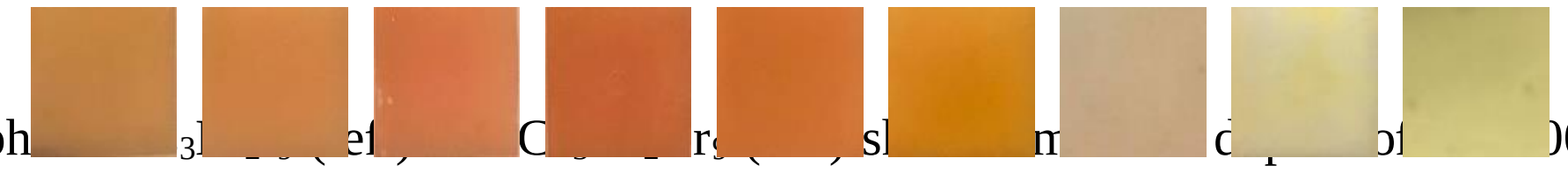

Figure S6 Photograph [illegible] 00% $SbBr_3$ from left to right

.

**Lead-free quaternary compounds:**

Bi and Sb based quaternary compounds were synthesized by mixing stoichiometric molar solution of 2:1:1 of AX (A=MA, Cs, X = Br, I), $B^{III}X_3$ (B = Bi, Sb, X = Br, I) and $B^{I}X$ (B = Ag, Cu, Na, X = Cl, Br, I) in mixed DMF and DMSO solvents. One-step spinning program was employed with 2000 rpm for 30s. There are only a few established deposition recipe for quaternary perovskite-inspired materials.[6,7]

Examples of thin-films successfully deposited are listed below:

| # | A | B | X |
|---|---|---|---|
| 1 | Cs | Ag-Bi | Br |
| 2 | Cs | Ag-Bi | Br |
| 3 | Cs | Ag-Sb | Br |
| 4 | Cs | Cu-Bi | I |
| 5 | Cs | Cu-Sb | I |
| 6 | Cs | Cu-Sb | I |
| 7 | Rb | Ag-Bi | I |
| 8 | Rb | Cu-Sb | I |

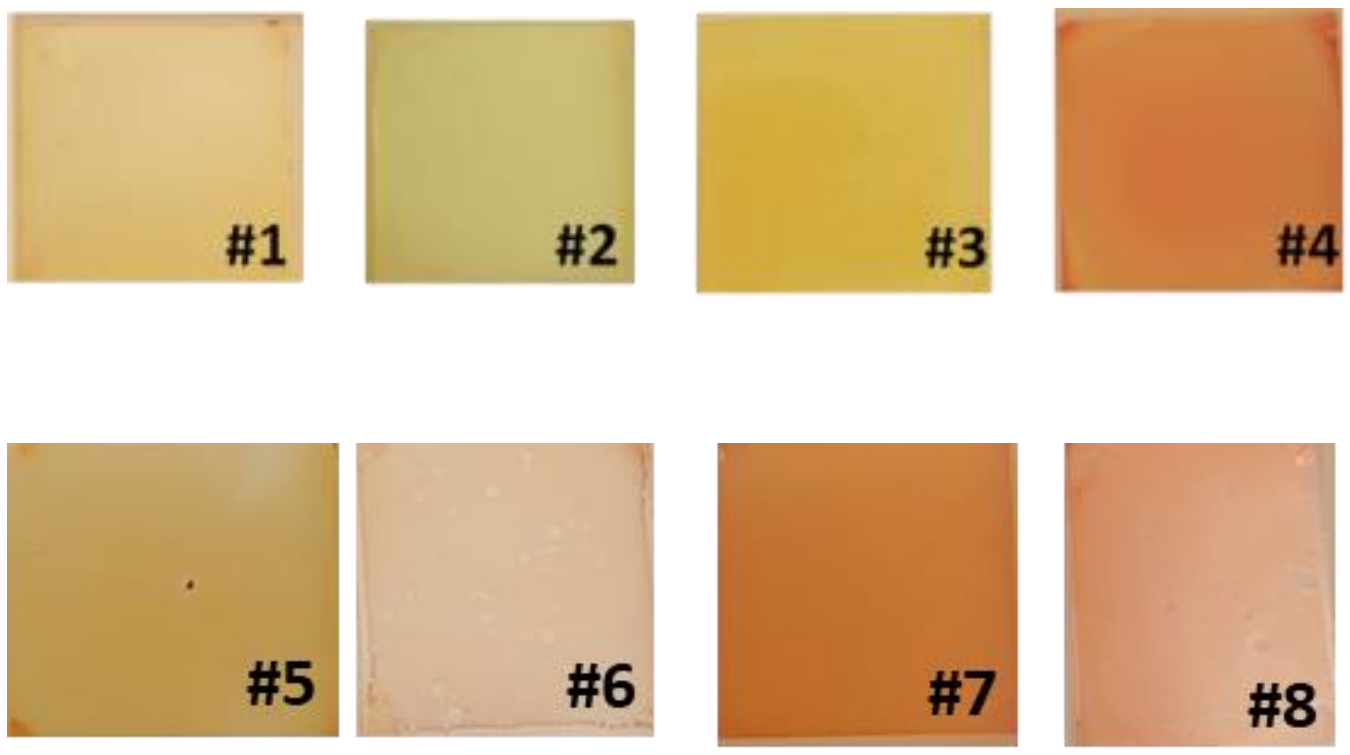

Table S2 Summary of 75 thin-film materials and their structural and optical properties. Raw data files (CSV) are available in a separate file.

| NO. | Compounds | | | | | Phase | | | | Bandgap from Tauc Plot (eV) | |
|---|---|---|---|---|---|---|---|---|---|---|---|
| Sample # | Target Compound | Recipe Ref. | Synthesized Compound | Materials Category | Made into films? (Y/N) | Dimen-sionality | Space Group | Other phases observed | Phase ID Ref. | Direct | Indirect |
| 1 | FAPbI3 | [8] | FAPbI3 | Pb compounds | Y | 3D | Pm-3m | | [8] | 1.52 | 1.48 |
| 2 | (MAFA)Pb(IBr)3 | [9] | FAPbI3:MAPbBr3 = 9:1 | Pb compounds | Y | 3D | Pm-3m | | [10] | 1.57 | 1.53 |
| 3 | (MAFA)Pb(IBr)3 | [9] | FAPbI3:MAPbBr3 = 8:2 | Pb compounds | Y | 3D | Pm-3m | | [10] | 1.61 | 1.52 |
| 4 | (MAFA)Pb(IBr)3 | [9] | FAPbI3:MAPbBr3 = 7:3 | Pb compounds | Y | 3D | Pm-3m | | [10] | 1.68 | 1.55 |
| 5 | (MAFA)Pb(IBr)3 | [9] | FAPbI3:MAPbBr3 = 6:4 | Pb compounds | Y | 3D | Pm-3m | | [10] | 1.75 | 1.69 |
| 6 | (MAFA)Pb(IBr)3 | [9] | FAPbI3:MAPbBr3 = 5:5 | Pb compounds | Y | 3D | Pm-3m | | [10] | 1.85 | 1.79 |
| 7 | (MAFA)Pb(IBr)3 | [9] | FAPbI3:MAPbBr3 = 4:6 | Pb compounds | Y | 3D | Pm-3m | | [10] | 1.93 | 1.87 |
| 8 | (MAFA)Pb(IBr)3 | [9] | FAPbI3:MAPbBr3 = 3:7 | Pb compounds | Y | 3D | Pm-3m | | [10] | 2.03 | 1.97 |
| 9 | (MAFA)Pb(IBr)3 | [9] | FAPbI3:MAPbBr3 = 2:8 | Pb compounds | Y | 3D | Pm-3m | | [10] | 2.09 | 2.04 |
| 10 | (MAFA)Pb(IBr)3 | [9] | FAPbI3:MAPbBr3 = 1:9 | Pb compounds | Y | 3D | Pm-3m | | [10] | 2.21 | 2.15 |
| 11 | MAPbBr3 | [9] | MAPbBr3 | Pb compounds | Y | 3D | Pm-3m | | [11] | 2.29 | 2.23 |
| 12 | (CsRbMAFA)Pb(IBr)3 | [2] | %5CsI, 5%RbI, FAPbI3:MAPbBr3 = 9:1 | Pb compounds | Y | 3D | Pm-3m | PbI2 | [10] | 1.59 | 1.52 |
| 13 | (CsRbMAFA)Pb(IBr)3 | [2] | %5CsI, 5%RbI, FAPbI3:MAPbBr3 = 5:1 | Pb compounds | Y | 3D | Pm-3m | PbI2 | [10] | 1.62 | 1.55 |
| 14 | (CsRbMAFA)Pb(IBr)3 | [2] | %5CsI, 5%RbI, FAPbI3:MAPbBr3 = 3:1 | Pb compounds | Y | 3D | Pm-3m | PbI2 | [10] | 1.68 | 1.63 |
| 15 | (CsRbMAFA)Pb(IBr)3 | [2] | %5CsI, 5%RbI, FAPbI3:MAPbBr3 = 2:1 | Pb compounds | Y | 3D | Pm-3m | PbI2 | [10] | 1.74 | 1.67 |
| 16 | (CsRbMAFA)Pb(IBr)3 | [2] | %5CsI, 5%RbI, FAPbI3:MAPbBr3 = 1:1 | Pb compounds | Y | 3D | Pm-3m | PbI2 | [10] | 1.88 | 1.76 |
| 17 | (CsRbMAFA)Pb(IBr)3 | [2] | %5CsI, 5%RbI, FAPbI3:MAPbBr3 = 1:2 | Pb compounds | Y | 3D | Pm-3m | PbI2 | [10] | 2.01 | 1.87 |

| | | | | | | | | | | | | |
|---|---|---|---|---|---|---|---|---|---|---|---|---|
| 18 | (CsRbMAFA)Pb(IBr)3 | [2] | %5CsI, 5%RbI, FAPbI3:MAPbBr3 = 1:3 | Pb compounds | Y | 3D | Pm-3m | PbI2 | [10] | 2.05 | 1.92 |
| 19 | (CsRbMAFA)Pb(IBr)3 | [2] | %5CsI, 5%RbI, FAPbI3:MAPbBr3 = 1:5 | Pb compounds | Y | 3D | Pm-3m | PbI2 | [10] | 2.13 | 1.96 |
| 20 | (CsRbMAFA)Pb(IBr)3 | [2] | %5CsI, 5%RbI, FAPbI3:MAPbBr3 = 1:9 | Pb compounds | Y | 3D | Pm-3m | PbI2 | [11] | 2.18 | 2.04 |
| 21 | FAPbBr3 | [2] | FAPbBr3 | Pb compounds | Y | 3D | Pm-3m | | [10] | 2.25 | 2.16 |
| 22 | MASnI3 | [12] | MASnI3 | Sn compounds | Y | 3D | I4/mcm | SnI4 | [10] | 1.3 | 1.15 |
| 23 | MASnCaI3 | this work | 1%CaI, MASnI3 | Sn compounds | Y | 3D | I4/mcm | SnI4 | [10] | 1.3 | 1.15 |
| 24 | MASnCaI3 | this work | 5%CaI, MASnI3 | Sn compounds | Y | 3D | I4/mcm | SnI4 | [10] | 1.33 | 1.15 |
| 25 | MASnCaI3 | this work | 10%CaI, MASnI3 | Sn compounds | Y | 3D | I4/mcm | SnI4 | [10] | 1.35 | 1.14 |
| 26 | MA(SnPb)(IBr)3 | this work | MASnI3: MAPbBr3 = 9:1 | Sn compounds | Y | 3D | Pm-3m | | [3] | 1.28 | 1.1 |
| 27 | MA(SnPb)(IBr)3 | this work | MASnI3: MAPbBr3 = 8:2 | Sn compounds | Y | 3D | Pm-3m | | [3] | 1.34 | 1.23 |
| 28 | MA(SnPb)(IBr)3 | this work | MASnI3: MAPbBr3 = 7:3 | Sn compounds | Y | 3D | Pm-3m | | [3] | 1.38 | 1.29 |
| 29 | MA(SnPb)(IBr)3 | this work | MASnI3: MAPbBr3 = 6:4 | Sn compounds | Y | 3D | Pm-3m | | [3] | 1.48 | 1.33 |
| 30 | MA(SnPb)(IBr)3 | this work | MASnI3: MAPbBr3 = 5:5 | Sn compounds | Y | 3D | Pm-3m | | [3] | 1.57 | 1.48 |
| 31 | MA(SnPb)(IBr)3 | this work | MASnI3: MAPbBr3 = 4:6 | Sn compounds | Y | 3D | Pm-3m | | [3] | 1.65 | 1.54 |
| 32 | MA(SnPb)(IBr)3 | this work | MASnI3: MAPbBr3 = 3:7 | Sn compounds | Y | 3D | Pm-3m | | [3] | 1.83 | 1.65 |
| 33 | MA(SnPb)(IBr)3 | this work | MASnI3: MAPbBr3 = 2:8 | Sn compounds | Y | 3D | Pm-3m | | [3] | 1.84 | 1.64 |
| 34 | MA(SnPb)(IBr)3 | this work | MASnI3: MAPbBr3 = 1:9 | Sn compounds | Y | 3D | Pm-3m | | [3] | 2.12 | 1.73 |
| 35 | Cs3Bi2I9 | [5] | Cs3Bi2I9 | Bi/Sb ternary compounds | Y | 0D | P63/mmc | | [13] | 2.3 | 2.05 |
| 36 | Cs3Sb2I9 | [4] | Cs3Sb2I9 | Bi/Sb ternary compounds | Y | 0D | P63/mmc | | [13] | 2.48 | 2.42 |
| 37 | Rb3Bi2I9 | [4] | Rb3Bi2I9 | Bi/Sb ternary compounds | Y | 2D | Pc | | [13] | 2.28 | 2.09 |
| 38 | Rb3Sb2I9 | [4] | Rb3Sb2I9 | Bi/Sb ternary compounds | Y | 2D | P2/n | | [13] | 2.18 | 1.98 |
| 39 | Cs3Bi2Br9 | this work | Cs3Bi2Br9 | Bi/Sb ternary compounds | Y | 2D | P-3m1 | | [13] | 2.73 | 2.61 |
| 40 | Cs3Sb2Br9 | this work | Cs3Sb2Br9 | Bi/Sb ternary compounds | Y | 2D | P-3m1 | | [13] | 2.58 | 2.44 |
| 41 | Rb3Bi2Br9 | this work | Rb3Bi2Br9 | Bi/Sb ternary compounds | Y | 2D | P-3m1 | | [13] | 2.71 | 2.65 |
| 42 | Rb3Sb2Br9 | this work | Rb3Sb2Br9 | Bi/Sb ternary compounds | Y | 2D | P-3m1 | | [13] | 2.74 | 2.85 |
| 43 | K3Bi2I9 | [4] | K3Bi2I9 | Bi/Sb ternary compounds | Y | 2D | P-3m1 | | [13] | 2.3 | 2.03 |
| 44 | K3Sb2I9 | [4] | K3Sb2I9 | Bi/Sb ternary compounds | Y | 2D | P-3m1 | | [13] | 2.28 | 2.08 |
| 45 | Cs3(BiSb)2I9 | this work | Cs3Bi2I9: Cs3Sb2I9= 9:1 | Bi/Sb ternary compounds | Y | 0D | P63/mmc | | [4] | 2.31 | 1.98 |
| 46 | Cs3(BiSb)2I9 | this work | Cs3Bi2I9: Cs3Sb2I9= 8:2 | Bi/Sb ternary compounds | Y | 0D | P63/mmc | | [4] | 2.31 | 1.97 |
| 47 | Cs3(BiSb)2I9 | this work | Cs3Bi2I9: Cs3Sb2I9= 7:3 | Bi/Sb ternary compounds | Y | 0D | P63/mmc | | [4] | 2.21 | 1.88 |
| 48 | Cs3(BiSb)2I9 | this work | Cs3Bi2I9: Cs3Sb2I9= 6:4 | Bi/Sb ternary compounds | Y | 0D | P63/mmc | | [4] | 2.21 | 1.87 |
| 49 | Cs3(BiSb)2I9 | this work | Cs3Bi2I9: Cs3Sb2I9= 5:5 | Bi/Sb ternary compounds | Y | 0D | P63/mmc | | [4] | 2.3 | 1.92 |
| 50 | Cs3(BiSb)2I9 | this work | Cs3Bi2I9: Cs3Sb2I9= 4:6 | Bi/Sb ternary compounds | Y | 0D | P63/mmc | | [4] | 2.31 | 1.94 |
| 51 | Cs3(BiSb)2I9 | this work | Cs3Bi2I9: Cs3Sb2I9= 3:7 | Bi/Sb ternary compounds | Y | 0D | P63/mmc | | [4] | 2.32 | 1.94 |
| 52 | Cs3(BiSb)2I9 | this work | Cs3Bi2I9: Cs3Sb2I9= 2:8 | Bi/Sb ternary compounds | Y | 0D | P63/mmc | | [4] | 2.46 | 1.99 |
| 53 | Cs3(BiSb)2I9 | this work | Cs3Bi2I9: Cs3Sb2I9= 1:9 | Bi/Sb ternary compounds | Y | 0D | P63/mmc | | [4] | 2.01 | 1.8 |
| 54 | Cs3(BiSb)2(IBr)9 | this work | Cs3Bi2I9: Cs3Sb2Br9 =9:1 | Bi/Sb ternary compounds | Y | 0D | P63/mmc | | this work | 2.27 | 2 |
| 55 | Cs3(BiSb)2(IBr)9 | this work | Cs3Bi2I9: Cs3Sb2Br9 =8:2 | Bi/Sb ternary compounds | Y | 2D | P-3m1 | | this work | 2.18 | 1.91 |
| 56 | Cs3(BiSb)2(IBr)9 | this work | Cs3Bi2I9: Cs3Sb2Br9 =7:3 | Bi/Sb ternary compounds | Y | 2D | P-3m1 | | this work | 2.18 | 1.95 |
| 57 | Cs3(BiSb)2(IBr)9 | this work | Cs3Bi2I9: Cs3Sb2Br9 =6:4 | Bi/Sb ternary compounds | Y | 2D | P-3m1 | | this work | 2.19 | 1.92 |
| 58 | Cs3(BiSb)2(IBr)9 | this work | Cs3Bi2I9: Cs3Sb2Br9 =5:5 | Bi/Sb ternary compounds | Y | 2D | P-3m1 | | this work | 2.23 | 1.97 |
| 59 | Cs3(BiSb)2(IBr)9 | this work | Cs3Bi2I9: Cs3Sb2Br9 =4:6 | Bi/Sb ternary compounds | Y | 2D | P-3m1 | | this work | 2.27 | 2.06 |
| 60 | Cs3(BiSb)2(IBr)9 | this work | Cs3Bi2I9: Cs3Sb2Br9 =3:7 | Bi/Sb ternary compounds | Y | 2D | P-3m1 | | this work | 2.39 | 2.11 |
| 61 | Cs3(BiSb)2(IBr)9 | this work | Cs3Bi2I9: Cs3Sb2Br9 =2:8 | Bi/Sb ternary compounds | Y | 2D | P-3m1 | | this work | 2.45 | 2.16 |
| 62 | Cs3(BiSb)2(IBr)9 | this work | Cs3Bi2I9: Cs3Sb2Br9 =1:9 | Bi/Sb ternary compounds | Y | 2D | P-3m1 | | this work | 2.48 | 2.17 |
| 63 | (MA)2CuBr2Cl2 | [14] | (MA)CuBr2Cl2 | Ag/Cu/Na-rich ternary and quaternary compounds | Y | 2D | Pnma | | [14] | 2.46 | 1.93 |
| 64 | (MA)CuCl4 | [14] | (MA)CuCl4 | Ag/Cu/Na-rich ternary and quaternary compounds | Y | 2D | Pnma | | [14] | 2.92 | 2.33 |
| 65 | Cs2AgBiBr6 | [15] | Cs2AgBiBr6 | Ag/Cu/Na-rich ternary and quaternary compounds | Y | 3D | Fm-3m | Cs3Bi2Br9 | [15] | 2.32 | 2.1 |
| 66 | Cs2AgSbBr6 | this work | Cs2AgSbBr6 | Ag/Cu/Na-rich ternary and quaternary compounds | Y | 3D (mix phases) | Fm-3m | Cs3Sb2Br9, Cs2SbBr6 | [16] | 2.27 | 1.89 |
| 67 | RbAgBiI | this work | Rb2AgBiI6 | Ag/Cu/Na-rich ternary and quaternary compounds | Y | 2D (mix phases) | P-3m1 | Rb3Bi2I9 | this work | 2.29 | 1.92 |
| 68 | CsNaBiI | this work | Cs2NaBiI6 | Ag/Cu/Na-rich ternary and quaternary compounds | Y | 0D (mix phases) | P63/mmc | Cs3Bi2I9 | this work | 2.31 | 2.03 |
| 69 | CsNaSbI | this work | Cs2NaSbI6 | Ag/Cu/Na-rich ternary and quaternary compounds | Y | 2D (mix phases) | P-3m1 | Cs3Sb2I9 | this work | 2.33 | 1.97 |
| 70 | RbNaSbI | this work | Rb2NaSbI6 | Ag/Cu/Na-rich ternary and quaternary compounds | Y | 2D (mix phases) | P-3m1 | Rb3Sb2I9 | this work | 2.32 | 1.94 |
| 71 | RbNaSbBr | this work | Rb2NaSbBr6 | Ag/Cu/Na-rich ternary and quaternary compounds | Y | 2D (mix phases) | P-3m1 | Rb3Sb2Br9 | this work | 2.09 | 1.73 |
| 72 | CsCuBiI | this work | Cs2CuBiI6 | Ag/Cu/Na-rich ternary and quaternary compounds | Y | 0D (mix phases) | P63/mmc | Cs3Bi2I9 | this work | 2.11 | 1.81 |
| 73 | CsCuSbI | this work | Cs2CuSbI6 | Ag/Cu/Na-rich ternary and quaternary compounds | Y | 0D (mix phases) | P63/mmc | Cs3Sb2I9 | this work | 2.2 | 1.67 |
| 74 | RbCuSbI | this work | Rb2CuSbI6 | Ag/Cu/Na-rich ternary and quaternary compounds | Y | 2D (mix phases) | P-3m1 | Rb3Sb2I9 | this work | 2.18 | 2.02 |
| 75 | CsNaSbIBr | this work | Cs2NaSb(IBr)6 | Ag/Cu/Na-rich ternary and quaternary compounds | Y | 2D (mix phases) | P-3m1 | Cs3Sb2Br9 | this work | 2.60 | 2.42 |

Table S3 Processing conditions of the 75 thin-film samples. Raw data files (CSV) are available in a separate file.

| | Precursors | | | | | | | | Processing | | |
|---|---|---|---|---|---|---|---|---|---|---|---|
| Sample # | AX alloy ratios | AX | Solvent | BX alloy ratios | BX | Solvent | AX:BX | Precursor Concentration/M | Preheat /°C | Annealing /°C | Annealing time/min |
| 1 | 1 | FAI | DMSO:DMF = 1:9 | 1 | PbI2 | DMSO:DMF = 1:9 | 1:1 | 1.2 | RT | 100 | 10 |
| 2 | 9:1 | FAI, MABr | DMSO:DMF = 1:9 | 9:1 | PbI2, PbBr2 | DMSO:DMF = 1:9 | 1:1 | 1.2 | RT | 100 | 10 |
| 3 | 8:2 | FAI, MABr | DMSO:DMF = 1:9 | 8:2 | PbI2, PbBr2 | DMSO:DMF = 1:9 | 1:1 | 1.2 | RT | 100 | 10 |
| 4 | 7:3 | FAI, MABr | DMSO:DMF = 1:9 | 7:3 | PbI2, PbBr2 | DMSO:DMF = 1:9 | 1:1 | 1.2 | RT | 100 | 10 |
| 5 | 6:4 | FAI, MABr | DMSO:DMF = 1:9 | 6:4 | PbI2, PbBr2 | DMSO:DMF = 1:9 | 1:1 | 1.2 | RT | 100 | 10 |
| 6 | 5:5 | FAI, MABr | DMSO:DMF = 1:9 | 5:5 | PbI2, PbBr2 | DMSO:DMF = 1:9 | 1:1 | 1.2 | RT | 100 | 10 |
| 7 | 4:6 | FAI, MABr | DMSO:DMF = 1:9 | 4:6 | PbI2, PbBr2 | DMSO:DMF = 1:9 | 1:1 | 1.2 | RT | 100 | 10 |
| 8 | 3:7 | FAI, MABr | DMSO:DMF = 1:9 | 3:7 | PbI2, PbBr2 | DMSO:DMF = 1:9 | 1:1 | 1.2 | RT | 100 | 10 |
| 9 | 2:8 | FAI, MABr | DMSO:DMF = 1:9 | 2:8 | PbI2, PbBr2 | DMSO:DMF = 1:9 | 1:1 | 1.2 | RT | 100 | 10 |
| 10 | 1:9 | FAI, MABr | DMSO:DMF = 1:9 | 1:9 | PbI2, PbBr2 | DMSO:DMF = 1:9 | 1:1 | 1.2 | RT | 100 | 10 |
| 11 | 1 | MABr | DMSO:DMF = 1:9 | 1 | PbBr2 | DMSO:DMF = 1:9 | 1:1 | 1.2 | RT | 100 | 10 |
| 12 | 9:1 +5%CsI + 5% RbI | FAI, MABr, CsI, RbI | DMSO:DMF = 1:4 | 9:1 | PbI2, PbBr2 | DMSO:DMF = 1:4 | 1:1 | 1.2 | RT | 100 | 10 |
| 13 | 5:1 +5%CsI + 5% RbI | FAI, MABr, CsI, RbI | DMSO:DMF = 1:4 | 5:1 | PbI2, PbBr3 | DMSO:DMF = 1:4 | 1:1 | 1.2 | RT | 100 | 10 |
| 14 | 3:1 +5%CsI + 5% RbI | FAI, MABr, CsI, RbI | DMSO:DMF = 1:4 | 3:1 | PbI2, PbBr4 | DMSO:DMF = 1:4 | 1:1 | 1.2 | RT | 100 | 10 |
| 15 | 2:1 +5%CsI + 5% RbI | FAI, MABr, CsI, RbI | DMSO:DMF = 1:4 | 2:1 | PbI2, PbBr5 | DMSO:DMF = 1:4 | 1:1 | 1.2 | RT | 100 | 10 |
| 16 | 1:1 +5%CsI + 5% RbI | FAI, MABr, CsI, RbI | DMSO:DMF = 1:4 | 1:1 | PbI2, PbBr6 | DMSO:DMF = 1:4 | 1:1 | 1.2 | RT | 100 | 10 |
| 17 | 1:2 +5%CsI + 5% RbI | FAI, MABr, CsI, RbI | DMSO:DMF = 1:4 | 1:2 | PbI2, PbBr7 | DMSO:DMF = 1:4 | 1:1 | 1.2 | RT | 100 | 10 |
| 18 | 1:3 +5%CsI + 5% RbI | FAI, MABr, CsI, RbI | DMSO:DMF = 1:4 | 1:3 | PbI2, PbBr8 | DMSO:DMF = 1:4 | 1:1 | 1.2 | RT | 100 | 10 |
| 19 | 1:5 +5%CsI + 5% RbI | FAI, MABr, CsI, RbI | DMSO:DMF = 1:4 | 1:5 | PbI2, PbBr9 | DMSO:DMF = 1:4 | 1:1 | 1.2 | RT | 100 | 10 |
| 20 | 1:9 +5%CsI + 5% RbI | FAI, MABr, CsI, RbI | DMSO:DMF = 1:4 | 1:9 | PbI2, PbBr10 | DMSO:DMF = 1:4 | 1:1 | 1.2 | RT | 100 | 10 |
| 21 | 1 | FABr | DMSO:DMF = 1:9 | 1 | PbBr2 | DMSO:DMF = 1:9 | 1:1 | 1.2 | RT | 110 | 10 |
| 22 | 1 | MAI | DMSO:DMF = 1:9 | 1 | SnI2 | DMSO:DMF = 1:9 | 1:1 | 1.2 | RT | 100 | 10 |
| 23 | 1%CaI | MAI, CaI | DMSO:DMF = 1:9 | 1 | SnI2 | DMSO:DMF = 1:9 | 1:1 | 1.2 | RT | 100 | 10 |
| 24 | 5%CaI | MAI, CaI | DMSO:DMF = 1:9 | 1 | SnI2 | DMSO:DMF = 1:9 | 1:1 | 1.2 | RT | 100 | 10 |
| 25 | 10%CaI | MAI, CaI | DMSO:DMF = 1:9 | 1 | SnI2 | DMSO:DMF = 1:9 | 1:1 | 1.2 | RT | 100 | 10 |
| 26 | 9:1 | MAI, MABr | DMSO:DMF = 1:9 | 9:1 | SnI2, PbBr2 | DMSO:DMF = 1:9 | 1:1 | 1.2 | RT | 100 | 10 |
| 27 | 8:2 | MAI, MABr | DMSO:DMF = 1:9 | 8:2 | SnI2, PbBr2 | DMSO:DMF = 1:9 | 1:1 | 1.2 | RT | 100 | 10 |
| 28 | 7:3 | MAI, MABr | DMSO:DMF = 1:9 | 7:3 | SnI2, PbBr2 | DMSO:DMF = 1:9 | 1:1 | 1.2 | RT | 100 | 10 |
| 29 | 6:4 | MAI, MABr | DMSO:DMF = 1:9 | 6:4 | SnI2, PbBr2 | DMSO:DMF = 1:9 | 1:1 | 1.2 | RT | 100 | 10 |
| 30 | 5:5 | MAI, MABr | DMSO:DMF = 1:9 | 5:5 | SnI2, PbBr2 | DMSO:DMF = 1:9 | 1:1 | 1.2 | RT | 100 | 10 |
| 31 | 4:6 | MAI, MABr | DMSO:DMF = 1:9 | 4:6 | SnI2, PbBr2 | DMSO:DMF = 1:9 | 1:1 | 1.2 | RT | 100 | 10 |
| 32 | 3:7 | MAI, MABr | DMSO:DMF = 1:9 | 3:7 | SnI2, PbBr2 | DMSO:DMF = 1:9 | 1:1 | 1.2 | RT | 100 | 10 |
| 33 | 2:8 | MAI, MABr | DMSO:DMF = 1:9 | 2:8 | SnI2, PbBr2 | DMSO:DMF = 1:9 | 1:1 | 1.2 | RT | 100 | 10 |
| 34 | 1:9 | MAI, MABr | DMSO:DMF = 1:9 | 1:9 | SnI2, PbBr2 | DMSO:DMF = 1:9 | 1:1 | 1.2 | RT | 100 | 10 |
| 35 | 1 | CsI | DMSO:DMF = 1:0 | 1 | BiI3 | DMSO:DMF = 1:0 | 3:2 | 0.4 | 75 | 125 | 15 |
| 36 | 1 | CsI | DMSO:DMF =0:1 | 1 | SbI3 | DMSO:DMF = 1:0 | 3:2 | 0.4 | 75 | 150 | 10 |
| 37 | 1 | RbI | DMSO:DMF =1:0 | 1 | BiI3 | DMSO:DMF = 1:0 | 3:2 | 0.4 | 75 | 150 | 10 |
| 38 | 1 | RbI | DMSO:DMF =0:1 | 1 | SbI3 | DMSO:DMF = 1:0 | 3:2 | 0.4 | 75 | 150 | 10 |
| 39 | 1 | CsBr | DMSO:DMF =1:0 | 1 | BiBr3 | DMSO:DMF = 1:0 | 3:2 | 0.4 | 75 | 150 | 30 |
| 40 | 1 | CsBr | DMSO:DMF =1:0 | 1 | SbBr3 | DMSO:DMF = 1:0 | 3:2 | 0.4 | 75 | 150 | 30 |
| 41 | 1 | RbBr | DMSO:DMF =1:0 | 1 | BiBr3 | DMSO:DMF = 1:0 | 3:2 | 0.4 | 75 | 150 | 30 |
| 42 | 1 | RbBr | DMSO:DMF =1:0 | 1 | SbBr3 | DMSO:DMF = 1:0 | 3:2 | 0.4 | 75 | 75 | 30 |
| 43 | 1 | KI | DMSO:DMF =1:0 | 1 | BiBI3 | DMSO:DMF = 1:0 | 3:2 | 0.4 | 75 | 150 | 30 |
| 44 | 1 | KI | DMSO:DMF =0:1 | 1 | SbI3 | DMSO:DMF = 1:0 | 3:2 | 0.4 | 75 | 150 | 30 |
| 45* | 1 | CsI | DMSO:DMF = 1:1 | 9:1 | BiI3, SbI3 | DMSO:DMF = 1:1 | 3:2 | 0.4 | 75 | 150 | 15 |
| 46 | 1 | CsI | DMSO:DMF = 1:1 | 8:2 | BiI3, SbI3 | DMSO:DMF = 1:1 | 3:2 | 0.4 | 75 | 150 | 15 |
| 47 | 1 | CsI | DMSO:DMF = 1:1 | 7:3 | BiI3, SbI3 | DMSO:DMF = 1:1 | 3:2 | 0.4 | 75 | 150 | 15 |
| 48 | 1 | CsI | DMSO:DMF = 1:1 | 6:4 | BiI3, SbI3 | DMSO:DMF = 1:1 | 3:2 | 0.4 | 75 | 150 | 15 |
| 49 | 1 | CsI | DMSO:DMF = 1:1 | 5:5 | BiI3, SbI3 | DMSO:DMF = 1:1 | 3:2 | 0.4 | 75 | 150 | 15 |
| 50 | 1 | CsI | DMSO:DMF = 1:1 | 4:6 | BiI3, SbI3 | DMSO:DMF = 1:1 | 3:2 | 0.4 | 75 | 150 | 15 |
| 51 | 1 | CsI | DMSO:DMF = 1:1 | 3:7 | BiI3, SbI3 | DMSO:DMF = 1:1 | 3:2 | 0.4 | 75 | 150 | 15 |
| 52 | 1 | CsI | DMSO:DMF = 1:1 | 2:8 | BiI3, SbI3 | DMSO:DMF = 1:1 | 3:2 | 0.4 | 75 | 150 | 15 |
| 53 | 1 | CsI | DMSO:DMF = 1:1 | 1:9 | BiI3, SbI3 | DMSO:DMF = 1:1 | 3:2 | 0.4 | 75 | 150 | 15 |
| 54 | 9:1 | CsI, CsBr | DMSO:DMF = 1:1 | 9:1 | BiI3, SbBr3 | DMSO:DMF = 1:1 | 3:2 | 0.38 | 75 | 125 | 15 |
| 55 | 8:2 | CsI, CsBr | DMSO:DMF = 1:1 | 8:2 | BiI3, SbBr3 | DMSO:DMF = 1:1 | 3:2 | 0.38 | 75 | 125 | 15 |

| | | | | | | | | | | | | |
|---|---|---|---|---|---|---|---|---|---|---|---|---|
| 56 | 7:3 | CsI, CsBr | DMSO:DMF = 1:1 | 7:3 | BiI3, SbBr3 | DMSO:DMF = 1:1 | 3:2 | 0.38 | 75 | 125 | 15 |
| 57 | 6:4 | CsI, CsBr | DMSO:DMF = 1:1 | 6:4 | BiI3, SbBr3 | DMSO:DMF = 1:1 | 3:2 | 0.38 | 75 | 125 | 15 |
| 58 | 5:5 | CsI, CsBr | DMSO:DMF = 1:1 | 5:5 | BiI3, SbBr3 | DMSO:DMF = 1:1 | 3:2 | 0.38 | 75 | 125 | 15 |
| 59 | 4:6 | CsI, CsBr | DMSO:DMF = 1:1 | 4:6 | BiI3, SbBr3 | DMSO:DMF = 1:1 | 3:2 | 0.38 | 75 | 125 | 15 |
| 60 | 3:7 | CsI, CsBr | DMSO:DMF = 1:1 | 3:7 | BiI3, SbBr3 | DMSO:DMF = 1:1 | 3:2 | 0.38 | 75 | 125 | 15 |
| 61 | 2:8 | CsI, CsBr | DMSO:DMF = 1:1 | 2:8 | BiI3, SbBr3 | DMSO:DMF = 1:1 | 3:2 | 0.38 | 75 | 125 | 15 |
| 62 | 1:9 | CsI, CsBr | DMSO:DMF = 1:1 | 1:9 | BiI3, SbBr3 | DMSO:DMF = 1:1 | 3:2 | 0.38 | 75 | 125 | 15 |
| 63 | 1 | MABr | DMSO:DMF = 0:1 | 1 | CuCl2 | DMSO:DMF = 1:0 | 2:1 | 0.6 | 75 | 80 | 10 |
| 64 | 1 | MACl | DMSO:DMF = 0:1 | 1 | CuCl2 | DMSO:DMF = 1:0 | 2:1 | 0.6 | 75 | 80 | 10 |
| 65 | 1 | CsBr | DMSO:DMF = 1:0 | 1:1 | AgBr, BiBr3 | DMSO:DMF = 1:0 | 2:1 | 0.6 | 75 | 285 | 5 |
| 66 | 1 | CsBr | DMSO:DMF = 1:0 | 1:1 | AgBr, SbBr3 | DMSO:DMF = 1:0 | 2:1 | 0.6 | 200 | 150 | 30 |
| 67 | 1 | RbI | Butylamine | 1:1 | AgI, BiI3 | DMSO:DMF = 1:1 | 2:1 | 0.6 | RT | 150 | 5 |
| 68 | 1 | CsI | DMSO:DMF = 1:0 | 1:1 | NaI, BiI3 | DMSO:DMF = 1:0 | 2:1 | 0.6 | 75 | 110 | 10 |
| 69 | 1 | CsI | DMSO:DMF = 1:0 | 1:1 | NaI, SbI3 | DMSO:DMF = 1:0 | 2:1 | 0.6 | 75 | 110 | 10 |
| 70 | 1 | RbI | DMSO:DMF = 1:0 | 1:1 | NaI, SbI3 | DMSO:DMF = 1:0 | 2:1 | 0.6 | 75 | 110 | 10 |
| 71 | 1 | RbBr | DMSO:DMF = 1:0 | 1:1 | NaBr, SbBr3 | DMSO:DMF = 1:0 | 2:1 | 0.6 | 75 | 110 | 10 |
| 72 | 1 | CsI | DMSO:DMF = 1:0 | 1:1 | CuI, BiI3 | DMSO:DMF = 0:1 | 2:1 | 0.6 | 75 | 285 | 5 |
| 73 | 1 | CsI | DMSO:DMF = 1:0 | 1:1 | CuI, SbI3 | DMSO:DMF = 0:1 | 2:1 | 0.6 | RT | 285 | 5 |
| 74 | 1 | RbI | DMSO:DMF = 1:0 | 1:1 | CuI, SbI3 | DMSO:DMF = 0:1 | 2:1 | 0.6 | 75 | 285 | 5 |
| 75 | 1:1 | CsI | DMSO:DMF = 1:0 | 1:1 | NaBr, SbBr | DMSO:DMF = 0:1 | 2:1 | 0.6 | RT | 110 | 10 |

Note: For sample no. 45-62, an optimized recipe consists of dissolving Cs$X$:Bi$X_3$ = 3:2 in DMSO and then the solution was mixed with Cs$X$:Sb$X_3$ = 3:2 ($X$ = I, Br) in DMF.

## III. Characterization

Transmission and reflection was measured for as-synthesized thin-film samples using Perkin-Elmer Lambda 950 UV/Vis Spectrophotometer. Absorptance was calculated using $A = 1\text{-}T\text{-}R$. Using the method established by Tauc,[17] we extracted the band gap for the thin films. Band gaps were calculated for both direct and indirect bandgap assumption. Approximate thicknesses of 300 nm film for 1.2M precursor concentration and 100nm film for 0.6M precursor concentration were used for optical properties estimation.

1.

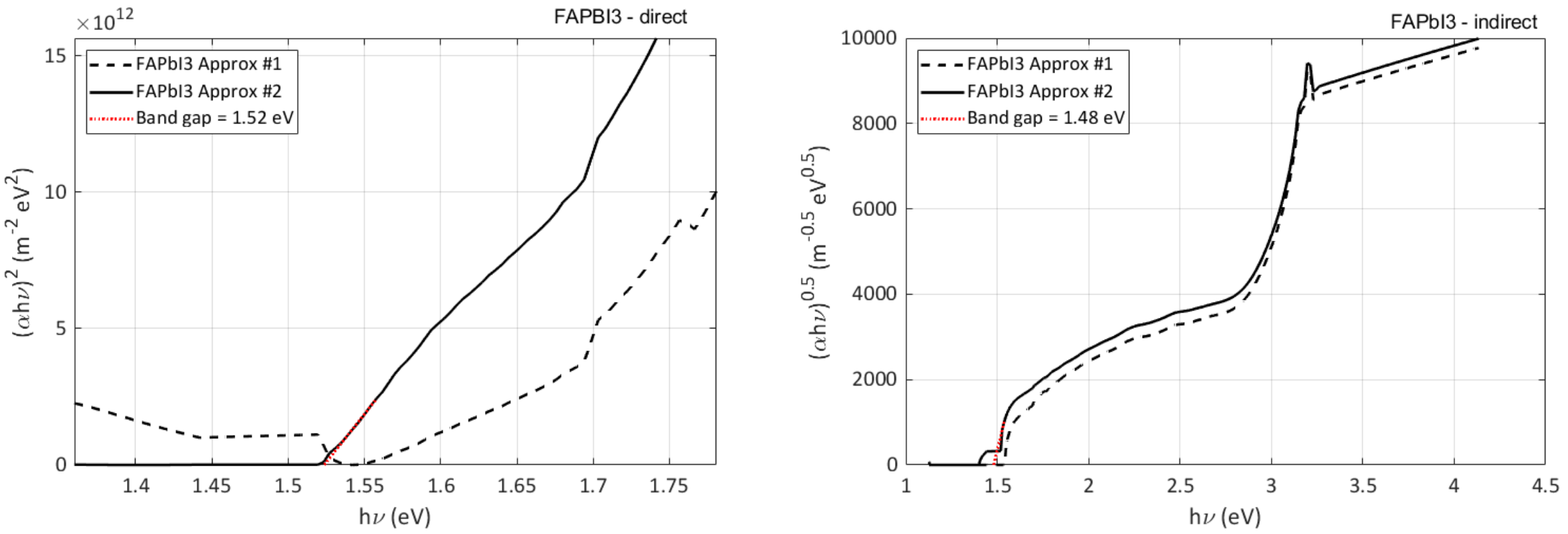

Figure S7 Tauc plots thin-films (Sample No.1), with direct and indirect bandgap assumptions. A lamp change was employed at 850 nm. Raw data files for Sample No. 1-75 are available in a separate file.

PXRD measurement was conducted using a Rigaku SmartLab diffractometer. Parallel beam geometry with a step size of 0.04° and 2θ range of 5°– 60° was employed for theta-omega scans of synthesized films. The $MA_{1-x}FA_xPbI_{3x}Br_{3-x}$, $MASn_{1-x}Ca_xI_3$ and $MASn_{1-x}Pb_xI_{3-x}Br_{3x}$ and $Cs_3Bi_{2-2x}Sb_{2x}I_{9x}Br_{9-9x}$, $MASn_{1-x}Ca_xI_3$ and $MASn_{1-x}Pb_xI_{3-x}Br_{3x}$ series were measured with grazing incidence PXRD measurement with the same step size, due to the highly orientated films with preferred orientation of {*h00*} in a cubic perovskite and {*00l*} in layered perovskites. Pawley refinement was carried out using Topas Academic V6 for structure refinement.[18]

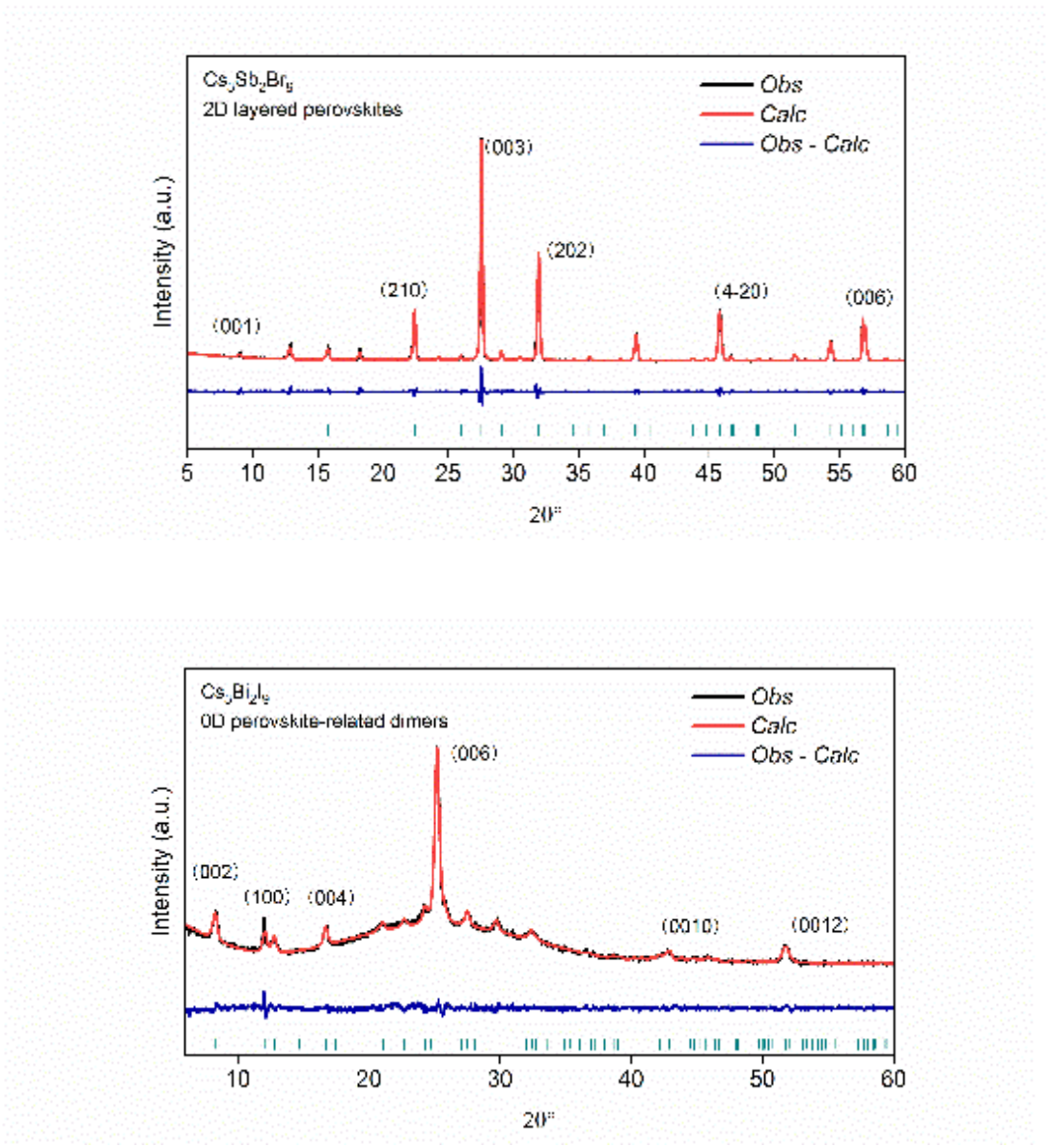

Figure S8 Pawley refinement of the 2D layered perovskite, $Cs_3Sb_2Br_9$ and the 0D dimer, $Cs_3BiI_9$. The alloy series of these two materials is presented in Figure 5.

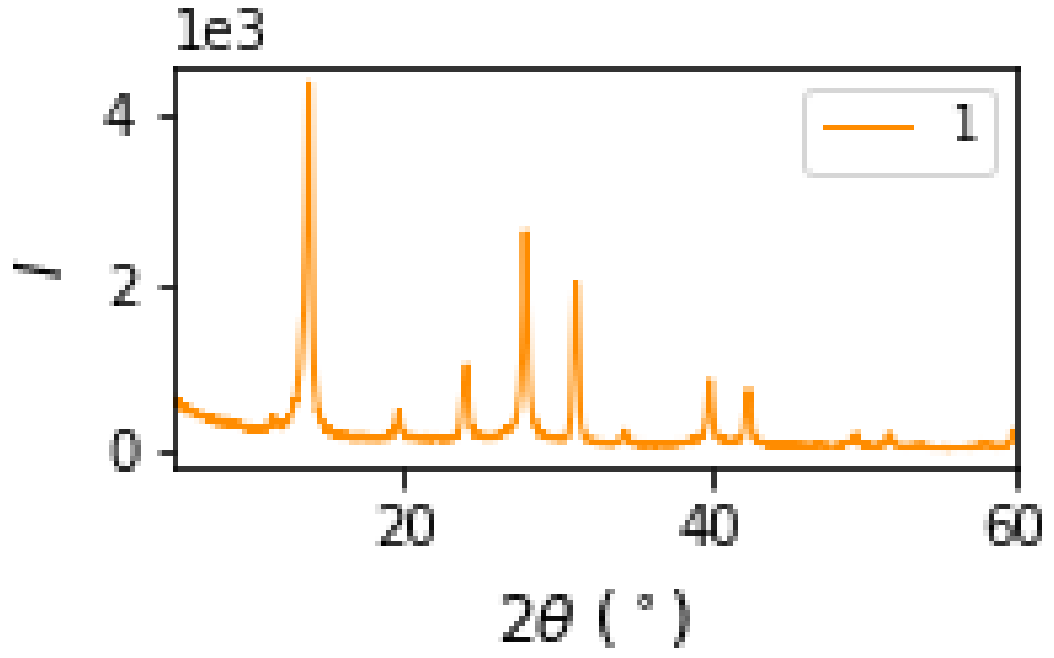

Figure S9 Experimental PXRD patterns of the deposited thin-film sample (Sample No.1) in Table S2. Raw data files for Sample No. 1-75 are available in a separate file.

# IV. Machine-learning methods

The simulated training dataset of XRD patterns for the machine-learning approach consists of 150 patterns extracted from compounds available in the Inorganic Crystal Structure Database (ICSD). Simulations of XRD powder patterns from the ICSD crystal structure information were carried out with Panalytical Highscore v4.7 software, based on the Rietveld algorithm implemented by Hill and Howard.[19] The empirical XRD patterns were preprocessed with a background subtraction and smoothed by a Savitzky Golay filter.

A number of classification algorithms were tested to determine the best performing algorithm using the simulated and experimental augmented datasets. The most accurate method was found to be a deep feedforward neural network of 3 layers composed of 256 neurons each. Stochastic gradient descent was used for optimization. The neural network was implemented using the vanilla algorithm for Multilayer Perceptron in ScikitLearn.[20]

Three approaches were taken with different training and test dataset:

1. The first approach involved using exclusively the simulated XRD dataset. After 5-fold cross validation on the simulated spectrum, a model accuracy of **99%** was estimated.
2. The second approach consisted in using the simulated XRD patterns as a training dataset, and the experimental patterns for known materials were used as a testing dataset. After cross validation, the model accuracy was **76%.**
3. The third approach consisted in using both experimental and simulated data for training. The full simulated dataset and 80% of the experimental known-material dataset were employed. Subsequently, 20% of the experimental dataset was left out for testing. After cross validation, the model accuracy of **90%** was estimated. This approach was then employed to test the novel materials.

The first approach has a much higher accuracy as it does not predict any experimental data and thus is free of experimental errors. The second approach does the experimental prediction solely based on simulated diffraction patterns. It has the lowest accuracy. The last approach has a significant higher accuracy than the second approach but lower than the first approach. The use of experimental data as part of the training set, increases the model accuracy and robustness.

To evaluate trade-offs between data quality and acquisition speed, we further investigated how the data coarsening will impact the accuracy of out prediction. We found that fast X-ray measurement could be achieved by increasing the step size, while still satisfy the requirement for accuracy. 90% accuracy achieved when the 2 theta step size is less than 0.16 degree. In this study, a 0.04 step size was used.

Table S4 List of the ML classification of the 75 thin-film materials, and their confidence score for each dimensionality. The sample IDs are following the compounds listed in Table S2.

| Data Labels | | | Confidence Score | | |
|---|---|---|---|---|---|
| Sample # | **Materials Category** | Group | 0D | 2D | 3D |
| 1 | Pb compounds | 1 | 0 | 0.018126 | 0.981874 |
| 2 | Pb compounds | 1 | 0.014662 | 0.021904 | 0.963434 |
| 3 | Pb compounds | 1 | 0 | 0.01117 | 0.98883 |
| 4 | Pb compounds | 1 | 0.006212 | 0.014673 | 0.979115 |
| 5 | Pb compounds | 1 | 0.003583 | 0.014295 | 0.982123 |
| 6 | Pb compounds | 1 | 0.001994 | 0 | 0.998006 |
| 7 | Pb compounds | 1 | 0.004627 | 0.001113 | 0.99426 |
| 8 | Pb compounds | 1 | 6.08E-04 | 0 | 0.999392 |
| 9 | Pb compounds | 1 | 0 | 0.013803 | 0.986197 |
| 10 | Pb compounds | 1 | 0 | 0.013323 | 0.986677 |
| 11 | Pb compounds | 1 | 0 | 0.012499 | 0.987501 |
| 12 | Pb compounds | 1 | 0 | 0.000552 | 0.999448 |
| 13 | Pb compounds | 1 | 0 | 1.02E-02 | 0.989815 |
| 14 | Pb compounds | 1 | 0.009113 | 0.005155 | 0.985733 |
| 15 | Pb compounds | 1 | 0.040542 | 0.031396 | 0.928062 |
| 16 | Pb compounds | 1 | 0 | 0.008727 | 0.991273 |
| 17 | Pb compounds | 1 | 0.004603 | 0.013025 | 0.982372 |
| 18 | Pb compounds | 1 | 0 | 0.001617 | 0.998383 |
| 19 | Pb compounds | 1 | 0 | 0.00976 | 0.99024 |
| 20 | Pb compounds | 1 | 0.039774 | 0 | 0.960226 |
| 21 | Pb compounds | 1 | 0 | 0.024624 | 0.975376 |
| 22 | Sn compounds | 1 | 0.008245 | 0 | 0.991755 |
| 23 | Sn compounds | 1 | 0.007111 | 0 | 0.992889 |
| 24 | Sn compounds | 1 | 0 | 0.000522 | 0.999478 |
| 25 | Sn compounds | 1 | 0.001068 | 0.007415 | 0.991516 |
| 26 | Sn compounds | 1 | 0.008528 | 0 | 0.991472 |
| 27 | Sn compounds | 1 | 0.001954 | 0 | 0.998046 |
| 28 | Sn compounds | 1 | 0.002876 | 0 | 0.997124 |
| 29 | Sn compounds | 1 | 0.005865 | 0.002657 | 0.991477 |
| 30 | Sn compounds | 1 | 0.064121 | 0 | 0.935879 |
| 31 | Sn compounds | 1 | 0.014971 | 0.014162 | 0.970868 |
| 32 | Sn compounds | 1 | 0 | 0.021745 | 0.978255 |
| 33 | Sn compounds | 1 | 0 | 0.020309 | 0.979691 |
| 34 | Sn compounds | 1 | 0.056232 | 0.01127 | 0.932498 |
| 35 | Bi/Sb ternary compounds | 1 | 1 | 0 | 0 |
| 36 | Bi/Sb ternary compounds | 1 | 0.98968 | 0.01032 | 0 |
| 37 | Bi/Sb ternary compounds | 1 | 0.178615 | 0.821385 | 0 |
| 38 | Bi/Sb ternary compounds | 1 | 0 | 1 | 0 |
| 39 | Bi/Sb ternary compounds | 1 | 0.00466 | 0.99534 | 0 |
| 40 | Bi/Sb ternary compounds | 1 | 0 | 1 | 0 |
| 41 | Bi/Sb ternary compounds | 1 | 0 | 0.985905 | 0.014095 |
| 42 | Bi/Sb ternary compounds | 1 | 0.003284 | 0.996716 | 0 |
| 43 | Bi/Sb ternary compounds | 1 | 0.046219 | 0.953781 | 0 |
| 44 | Bi/Sb ternary compounds | 1 | 0.02417 | 0.97583 | 0 |
| 45 | Bi/Sb ternary compounds | 1 | 0.808698 | 0.191302 | 0 |
| 46 | Bi/Sb ternary compounds | 1 | 0.970068 | 0.029932 | 0 |
| 47 | Bi/Sb ternary compounds | 1 | 1 | 0 | 0 |
| 48 | Bi/Sb ternary compounds | 1 | 1 | 0 | 0 |
| 49 | Bi/Sb ternary compounds | 1 | 1 | 0 | 0 |
| 50 | Bi/Sb ternary compounds | 1 | 1 | 0 | 0 |
| 51 | Bi/Sb ternary compounds | 1 | 1 | 0 | 0 |
| 52 | Bi/Sb ternary compounds | 1 | 1 | 0 | 0 |
| 53 | Bi/Sb ternary compounds | 1 | 0.960211 | 0.035819 | 0.00397 |
| 54 | Bi/Sb ternary compounds | 2 | 0.830799 | 0.169201 | 0 |
| 55 | Bi/Sb ternary compounds | 2 | 0.04218 | 0.95782 | 0 |
| 56 | Bi/Sb ternary compounds | 2 | 0 | 1 | 0 |
| 57 | Bi/Sb ternary compounds | 2 | 0 | 0.986883 | 0.013117 |
| 58 | Bi/Sb ternary compounds | 2 | 0 | 0.9654 | 0.0346 |
| 59 | Bi/Sb ternary compounds | 2 | 0.015704 | 0.946217 | 0.038079 |
| 60 | Bi/Sb ternary compounds | 2 | 0.015526 | 0.886036 | 0.098438 |
| 61 | Bi/Sb ternary compounds | 2 | 0 | 0.962374 | 0.037626 |
| 62 | Bi/Sb ternary compounds | 2 | 0 | 0.965981 | 0.034019 |
| 63 | Ag/Cu/Na ternary and quaternary compounds | 2 | Pnma symmetry, Not included | | |
| 64 | Ag/Cu/Na ternary and quaternary compounds | 2 | Pnma symmetry, Not included | | |
| 65 | Ag/Cu/Na ternary and quaternary compounds | 1 | 0.017502 | 0 | 0.982498 |
| 66 | Ag/Cu/Na ternary and quaternary compounds | 1 | 0.003225 | 0 | 0.996775 |
| 67 | Ag/Cu/Na ternary and quaternary compounds | 2 | 0 | 0.031641 | 0.968359 |
| 68 | Ag/Cu/Na ternary and quaternary compounds | 2 | 0.441636 | 0.558364 | 0 |
| 69 | Ag/Cu/Na ternary and quaternary compounds | 2 | 0.058699 | 0.894531 | 0.04677 |
| 70 | Ag/Cu/Na ternary and quaternary compounds | 2 | 0 | 0.996315 | 0.003685 |
| 71 | Ag/Cu/Na ternary and quaternary compounds | 2 | 0.002025 | 0.99725 | 0.000725 |
| 72 | Ag/Cu/Na ternary and quaternary compounds | 2 | 0.669369 | 0.330631 | 0 |
| 73 | Ag/Cu/Na ternary and quaternary compounds | 2 | 0.480775 | 0.519225 | 0 |
| 74 | Ag/Cu/Na ternary and quaternary compounds | 2 | 0.005867 | 0.994133 | 0 |
| 75 | Ag/Cu/Na ternary and quaternary compounds | 2 | 0.00052 | 0.961302 | 0.038178 |

# IV Additional experimental details

Table S5 List of unsuccessful depositions (Sample ID 76-96).

| Sample # | Target Compound | Materials Category | Why discarded? |
|---|---|---|---|
| 76 | (MA)2CuI4 | Ag/Cu/Na-rich ternary and quaternary compounds | Mixture deposited not identifiable |
| 77 | Na3Bi2Br9 | Bi/Sb ternary compounds | Mixture deposited not identifiable |
| 78 | Na3Sb2Br9 | Bi/Sb ternary compounds | Mixture deposited not identifiable |
| 79 | MA3Sb(IBr)9 | Bi/Sb ternary compounds | Mixture deposited not identifiable |
| 80 | (MA)3Sb2(ICl)9 | Bi/Sb ternary compounds | Mixture deposited not identifiable |
| 81 | Rb2AgSbI6 | Ag/Cu/Na-rich ternary and quaternary compounds | Not soluble |
| 82 | Cs2BiAgBr6 (Cs excess) | Ag/Cu/Na-rich ternary and quaternary compounds | Not soluble |
| 83 | Cs2AgBiI6 | Ag/Cu/Na-rich ternary and quaternary compounds | Not soluble |
| 84 | Cs2AgSbI6 | Ag/Cu/Na-rich ternary and quaternary compounds | Not soluble |
| 85 | Cs2NaBiBr6 | Ag/Cu/Na-rich ternary and quaternary compounds | Not soluble |
| 86 | Cs4(CuSb)2Cl12 | Ag/Cu/Na-rich ternary and quaternary compounds | Not soluble |
| 87 | Rb4(CuSb)2Cl12 | Ag/Cu/Na-rich ternary and quaternary compounds | Not soluble |
| 88 | MA3(CuSb)2l12 | Ag/Cu/Na-rich ternary and quaternary compounds | Not soluble |
| 89 | MA4(CuSb)2Cl12 | Ag/Cu/Na-rich ternary and quaternary compounds | Not soluble |
| 90 | MA4(CuSb)2Br12 | Ag/Cu/Na-rich ternary and quaternary compounds | Not soluble |
| 92 | CsNASbBr | Ag/Cu/Na-rich ternary and quaternary compounds | Not soluble |
| 93 | Cs(NaSbBi)I6 | Ag/Cu/Na-rich ternary and quaternary compounds | Not soluble |
| 94 | (CsRb)2Na(BiSb)(IBr)6 | Ag/Cu/Na-rich ternary and quaternary compounds | Mixture deposited not identifiable |
| 95 | Rb2(BiSb)(IBr)6 | Ag/Cu/Na-rich ternary and quaternary compounds | Mixture deposited not identifiable |
| 96 | RbNa(SbBi)I6 | Ag/Cu/Na-rich ternary and quaternary compounds | Mixture deposited not identifiable |

Table S6 Extended experimental notes and synthesis of additional offline repeated synthesis.

| Materials | | Bandgap from Tauc Plot (eV) | | Precursors | | | | Processing | | |
|---|---|---|---|---|---|---|---|---|---|---|
| Synthesized Compound | Materials Category | Direct | Indirect | AX | Solvent | BX | Precursor Concentration/ M | Preheat | Annealing | Annealing time/min |
| FAPbI3:MAPbBr3 = 9:1 | Pb compounds | 1.59 | 1.52 | FAI, MABr | DMSO: DMF = 1:9 | PbI2, PbBr2 | 1.2 | RT | 100 | 10 |
| FAPbI3:MAPbBr3 = 5:5 | Pb compounds | 1.89 | 1.83 | FAI, MABr | DMSO: DMF = 1:9 | PbI2, PbBr2 | 1.2 | RT | 100 | 10 |
| FAPbI3:MAPbBr3 = 1:9 | Pb compounds | 2.22 | 2.13 | FAI, MABr | DMSO: DMF = 1:9 | PbI2, PbBr2 | 1.2 | RT | 100 | 10 |
| MAPbBr3 | Pb compounds | 2.29 | 2.21 | MABr | DMSO: DMF = 1:9 | PbBr2 | 1.2 | RT | 100 | 10 |
| MAPbBr3 | Pb compounds | 2.42 | 2.21 | MABr | DMSO: DMF = 1:9 | PbBr2 | 1.2 | RT | 100 | 10 |
| MAPbBr3 | Pb compounds | 2.3 | 2.25 | MABr | DMSO: DMF = 1:9 | PbBr2 | 1.2 | RT | 100 | 10 |
| MASnI3 | Sn compounds | 1.33 | 1.15 | MAI | DMSO: DMF = 1:9 | SnI2 | 1.5 | RT | 100 | 10 |
| Cs3Bi2I9 | Bi/Sb ternary compounds | 2.29 | 2.03 | CsAc, | DMSO: DMF = 1:0 | BiI3 | 1 | 75 | 150 | 10 |
| Cs3Bi2I9 | Bi/Sb ternary compounds | 2.3 | 2.05 | CsI | DMSO: DMF = 1:0 | BiI3 | 0.4 | 75 | 150 | 10 |
| Cs3Bi2I9 | Bi/Sb ternary compounds | 2.63 | 2.16 | CsAc | DMSO: DMF = 1:0 | BiI3 | 0.4 | 75 | 150 | 15 |
| Cs3Bi2I9 | Bi/Sb ternary compounds | 2.33 | 2.11 | CsI | DMSO: DMF = 1:0 | BiI3 | 0.4 | 75 | 150 | 15 |
| Cs3Bi2I9 | Bi/Sb ternary compounds | 2.32 | 1.99 | CsI | DMSO: DMF = 1:0 | BiI3 | 0.4 | 75 | 150 | 15 |
| Cs3Bi2I9 | Bi/Sb ternary compounds | 2.34 | 2.17 | CsAc | DMSO: DMF = 1:0 | BiI3 | 0.4 | RT | 125 | 15 |
| Cs3Bi2I9 | Bi/Sb ternary compounds | 2.27 | 2 | CsAc | DMSO: DMF = 1:0 | BiI3 | 0.4 | RT | 125 | 15 |
| Cs3Bi2I9 | Bi/Sb ternary compounds | 2.3 | 1.98 | CsI | DMSO: DMF = 1:0 | BiI3 | 0.4 | RT | 125 | 15 |
| Cs3Bi2I9 | Bi/Sb ternary compounds | 2.28 | 2.01 | CsI | DMSO: DMF = 1:0 | BiI3 | 0.4 | RT | 125 | 15 |
| Cs3Sb2I9 | Bi/Sb ternary compounds | 2.27 | 1.8 | CsAc | DMSO: DMF =0:1 | SbI3 | 0.4 | 75 | 100 | 15 |
| Cs3Sb2I9 | Bi/Sb ternary compounds | 2.07 | 1.83 | CsI | DMSO: DMF =0:1 | SbI3 | 0.4 | 75 | 100 | 15 |
| Cs3Sb2I9 | Bi/Sb ternary compounds | 2.08 | 1.85 | CsI | DMSO: DMF =0:1 | SbI3 | 0.4 | 75 | 100 | 15 |
| Cs3Bi2Br9 | Bi/Sb ternary compounds | 3.05 | 2.86 | CsAc | DMSO: DMF =1:0 | BiBr3 | 0.4 | 75 | 150 | 15 |
| Cs3Bi2Br9 | Bi/Sb ternary compounds | 2.63 | 2.55 | CsBr | DMSO: DMF =1:0 | BiBr3 | 0.4 | 75 | 150 | 15 |
| Cs3Sb2Br9 | Bi/Sb ternary compounds | 3.44 | 2.44 | CsAc | DMSO: DMF =1:0 | SbBr3 | 1 | 75 | 150 | 10 |
| Cs3Sb2Br9 | Bi/Sb ternary compounds | 2.96 | 2.5 | CsAc | DMSO: DMF =1:0 | SbBr3 | 0.4 | 75 | 100 | 15 |
| Cs3Sb2Br9 | Bi/Sb ternary compounds | 3.04 | 2.64 | CsBr | DMSO: DMF =0:1 | SbBr3 | 0.4 | 75 | 100 | 15 |
| Cs3Sb2Br9 | Bi/Sb ternary compounds | 2.56 | 2.44 | CsAc | DMSO: DMF =1:0 | SbBr3 | 0.4 | RT | 125 | 15 |
| Cs3Sb2Br9 | Bi/Sb ternary compounds | 2.56 | 2.45 | CsAc | DMSO: DMF =1:0 | SbBr3 | 0.4 | 75 | 125 | 15 |
| Cs3Sb2Br9 | Bi/Sb ternary compounds | 2.63 | 2.43 | CsAc | DMSO: DMF =1:0 | SbBr3 | 0.4 | RT | 125 | 15 |
| Cs3Sb2Br9 | Bi/Sb ternary compounds | 2.63 | 2.47 | CsAc | DMSO: DMF =1:0 | SbBr3 | 0.4 | 75 | 125 | 15 |
| Cs3Sb2Br9 | Bi/Sb ternary compounds | 2.61 | 2.45 | CsBr | DMSO: DMF =1:0 | SbBr3 | 0.4 | RT | 125 | 15 |
| Cs3Sb2Br9 | Bi/Sb ternary compounds | 2.58 | 2.38 | CsBr | DMSO: DMF =1:0 | SbBr3 | 0.4 | 75 | 125 | 15 |
| Cs3Sb2Br9 | Bi/Sb ternary compounds | 2.64 | 2.48 | CsBr | DMSO: DMF =1:0 | SbBr3 | 0.4 | RT | 125 | 15 |
| Cs3Sb2Br9 | Bi/Sb ternary compounds | 2.64 | 2.51 | CsBr | DMSO: DMF =1:0 | SbBr3 | 0.4 | 75 | 125 | 15 |
| Cs3Sb2Br9 | Bi/Sb ternary compounds | 2.49 | 2.46 | CsBr | DMSO: DMF =1:0 | SbBr3 | 0.4 | 75 | 125 | 15 |
| Rb3Sb2Br9 | Bi/Sb ternary compounds | 3.08 | 2.85 | RbBr | DMSO: DMF =1:0 | SbBr3 | 0.4 | 75 | 150 | 30 |
| Cs3Bi2I9:Cs3Sb2Br9=9:1 | Bi/Sb ternary compounds | 2.36 | 2 | CsAc, CsAc | DMSO: DMF = 1:0 | BiI3, SbBr3 | 1 | 75 | 150 | 10 |
| Cs3Bi2I9:Cs3Sb2Br9=9:1 | Bi/Sb ternary compounds | 2.56 | 2.03 | CsAc, CsAc | DMSO: DMF = 1:0 | BiI3, SbBr3 | 0.4 | 75 | 150 | 15 |
| Cs3Bi2I9:Cs3Sb2Br9=8:2 | Bi/Sb ternary compounds | 2.27 | 1.82 | CsAc, CsAc | DMSO: DMF = 1:0 | BiI3, SbBr3 | 1 | 75 | 150 | 10 |
| Cs3Bi2I9:Cs3Sb2Br9=8:2 | Bi/Sb ternary compounds | 2.57 | 1.95 | CsAc, CsAc | DMSO: DMF = 1:0 | BiI3, SbBr3 | 0.4 | 75 | 150 | 15 |
| Cs3Bi2I9:Cs3Sb2Br9=8:2 | Bi/Sb ternary compounds | 2.21 | 1.92 | CsI, CsBr | DMSO: DMF = 1:1 | BiI3, SbBr3 | 0.4 | 75 | 150 | 15 |
| Cs3Bi2I9:Cs3Sb2Br9=7:3 | Bi/Sb ternary compounds | 2.27 | 1.94 | CsI, CsBr | DMSO: DMF = 1:1 | BiI3, SbBr3 | 0.4 | 75 | 150 | 15 |
| Cs3Bi2I9:Cs3Sb2Br9=6:4 | Bi/Sb ternary compounds | 2.25 | 2.01 | CsAc, CsAc | DMSO: DMF = 1:0 | BiI3, SbBr3 | 1 | 75 | 150 | 10 |
| Cs3Bi2I9:Cs3Sb2Br9=6:4 | Bi/Sb ternary compounds | 2.4 | 1.97 | CsAc, CsAc | DMSO: DMF = 1:0 | BiI3, SbBr3 | 0.4 | 75 | 150 | 15 |
| Cs3Bi2I9:Cs3Sb2Br9=6:4 | Bi/Sb ternary compounds | 2.28 | 1.95 | CsI, CsBr | DMSO: DMF = 1:1 | BiI3, SbBr3 | 0.4 | 75 | 150 | 15 |
| Cs3Bi2I9:Cs3Sb2Br9=5:5 | Bi/Sb ternary compounds | 2.29 | 2.08 | CsAc, CsAc | DMSO: DMF = 1:0 | BiI3, SbBr3 | 1 | 75 | 150 | 10 |
| Cs3Bi2I9:Cs3Sb2Br9=5:5 | Bi/Sb ternary compounds | 2.45 | 2.23 | CsAc, CsAc | DMSO: DMF = 1:0 | BiI3, SbBr3 | 0.4 | 75 | 150 | 15 |
| Cs3Bi2I9:Cs3Sb2Br9=5:5 | Bi/Sb ternary compounds | 2.27 | 2.03 | CsI, CsBr | DMSO: DMF = 1:1 | BiI3, SbBr3 | 0.4 | 75 | 150 | 15 |
| Cs3Bi2I9:Cs3Sb2Br9=4:6 | Bi/Sb ternary compounds | 2.34 | 2.02 | CsAc, CsAc | DMSO: DMF = 1:0 | BiI3, SbBr3 | 1 | 75 | 150 | 10 |
| Cs3Bi2I9:Cs3Sb2Br9=4:6 | Bi/Sb ternary compounds | 2.59 | 2.36 | CsAc, CsAc | DMSO: DMF = 1:0 | BiI3, SbBr3 | 0.4 | 75 | 150 | 15 |
| Cs3Bi2I9:Cs3Sb2Br9=4:6 | Bi/Sb ternary compounds | 2.29 | 1.95 | CsI, CsBr | DMSO: DMF = 1:1 | BiI3, SbBr3 | 0.4 | 75 | 150 | 15 |
| Cs3Bi2I9:Cs3Sb2Br9=3:7 | Bi/Sb ternary compounds | 2.37 | 2.21 | CsI, CsBr | DMSO: DMF = 1:1 | BiI3, SbBr3 | 0.4 | 75 | 150 | 15 |
| Cs3Bi2I9:Cs3Sb2Br9=2:8 | Bi/Sb ternary compounds | 2.44 | 2.18 | CsAc, CsAc | DMSO: DMF = 1:0 | BiI3, SbBr3 | 1 | 75 | 150 | 10 |
| Cs3Bi2I9:Cs3Sb2Br9=2:8 | Bi/Sb ternary compounds | 2.68 | 1.89 | CsAc, CsAc | DMSO: DMF = 1:0 | BiI3, SbBr3 | 0.4 | 75 | 150 | 15 |

| | | | | | | | | | |
|---|---|---|---|---|---|---|---|---|---|
| Cs3Bi2I9:Cs3Sb2Br9=2:8 | Bi/Sb ternary compounds | 2.41 | 2.11 | CsI, CsBr | DMSO: DMF = 1:1 | BiI3, SbBr3 | 0.4 | 75 | 150 | 15 |
| Cs3Bi2I9:Cs3Sb2Br9=1:9 | Bi/Sb ternary compounds | 2.54 | 2.27 | CsAc, CsAc | DMSO: DMF = 1:0 | BiI3, SbBr3 | 1 | 75 | 150 | 10 |
| Cs3Bi2I9:Cs3Sb2Br9=1:9 | Bi/Sb ternary compounds | 2.82 | 2.02 | CsAc, CsAc | DMSO: DMF = 1:0 | BiI3, SbBr3 | 0.4 | 75 | 150 | 15 |
| Cs3Bi2I9:Cs3Sb2Br9=1:9 | Bi/Sb ternary compounds | 2.63 | 2.29 | CsI, CsBr | DMSO: DMF = 1:1 | BiI3, SbBr3 | 0.4 | 75 | 150 | 15 |
| (MA)CuBr2Cl2 | Ag/Cu/Na-rich ternary and quartanary compounds | 2.38 | 1.89 | MABr | DMSO: DMF = 0:1 | CuCl2 | 0.6 | 75 | 80 | 10 |
| (MA)CuCl4 | Ag/Cu/Na-rich ternary and quartanary compounds | 2.9 | 2.4 | MACl | DMSO: DMF = 0:1 | CuCl2 | 0.6 | 75 | 80 | 10 |